# D-P2P-Sim+: A Novel Distributed Framework for P2P Protocols' Performance Testing*

*S. Sioutas*[1], *E. Sakkopoulos*[2], *A. Panaretos*[1], *D. Tsoumakos*[1], *P. Gerolymatos*[1],

*G. Tzimas*[4] *and Y. Manolopoulos*[3]

[1]Ionian University; [2]University of Patras; [3]Aristotle University of Thessaloniki; [4]Technological Educational Institute of Western Greece

## Abstract

In recent technologies like IoT (Internet of Things) and Web 2.0, a critical problem arises with respect to storing and processing the large amount of collected data. In this paper we develop and evaluate distributed infrastructures for storing and processing large amount of such data. We present a distributed framework that supports customized deployment of a variety of indexing engines over million-node overlays. The proposed framework provides the appropriate integrated set of tools that allows applications processing large amount of data, to evaluate and test the performance of various application protocols for very large scale deployments (multi million nodes - billions of keys). The key aim is to provide the appropriate environment that contributes in taking decisions regarding the choice of the protocol in storage P2P systems for a variety of big data applications. Using lightweight and efficient collection mechanisms, our system enables real-time registration of multiple measures, integrating support for real-life parameters such as node failure models and recovery strategies. Experiments have been performed at the PlanetLab network and at a typical research laboratory in order to verify scalability and show maximum re-usability of our setup. D-P2P-Sim+ framework is publicly available at http://code.google.com/p/d-p2p-sim/downloads/list.

* A very preliminary version of this work was presented in ACM SAC'12, pp.853-858, March 25-29, 2012, Riva del Garda, Italy. This work was partially supported by Thales Project entitled "Cloud9: A multidisciplinary, holistic approach to internet-scale cloud computing". For more details see the following URL: https://sites.google.com/site/thaliscloud9/home

# Introduction

Our era can be characterized by an explosion in the production of data: Internet of Things (IoT) and Web 2.0 technologies contribute to the production of increasing volumes of information that need to be stored, indexed and queried.

The Internet of things is a vision of the future Internet that expands beyond the virtual world to the physical word, through devices and wireless networks that enable the ubiquitous collection and transmission of information from uniquely identifiable objects. In (Atzori, Iera, & Morabito, 2010) the authors explain that the Internet of Things (IoT) cannot be restricted to just the application of RFIDs technology instead, it is a wider vision where devices, network and service technologies are combined introducing a new era of ubiquity. Devices with autonomous and proactive behavior, context awareness, capable of collaborative communications will contribute to IoT. In this plethora of information sources, scalability and personalization are necessary to handle massive heterogeneous resources as well as to enhance user-experience and allow access to the most relevant parts of the data. With respect to personalization according to (*Personalization reports*, n.d.) over 80% of users prefer the numerous personalization services that businesses and social sites offer.

Several problems have been identified in the literature that has to be addressed in order to realize the vision of IoT. One of them is scalability. As mentioned in (Miorandi, Sicari, De Pellegrini, & Chlamtac, 2012) the management of information and knowledge is creating scalability problems because for every entity in the physical world a virtual entity has to be built, moreover, massive services execution is required that need to handle massive heterogeneous resources. For example, RFIDs technology is being introduced in Supply Chain Management, to increase visibility, data quality and efficiency and better satisfy customers. As a result huge number of data is being generated. These large amounts of distributed data have to be exploited in a constructive way, they need for example to be analyzed and matched to the organization strategy. Constant monitoring of goods, real-time sales information collection and processing can benefit Supply Chain Management, allowing taking action in time and decreasing costs. Building a system to store and process this information adequately is challenging technically and financially (Zhang, Goh, & Meng, 2011).

Consider for example an infrastructure for storing and processing data collected from sensors like RFIDs that participate in the supply chain. Each member of the supply chain is interested in asking different questions in different sets of data. Personalization is utilized in IoT Supply Chain Management environment in different ways. A participant may play different roles based on the location and the device used hence different participant profiles are necessary. Managing digital identities and user profiles in IoT environments is problem that needs to be addressed in a large scale.

The sheer size of both data and their diverse sources render centralized, legacy systems inept at storing and processing them. The desired scalability and efficacy to handle the advanced processing required are provided by distributed solutions as both academic and business innovation has already indicated. Recently P2P storage solutions have been adopted to address these issues.

Designing new protocols and overlays for P2P systems includes apart from the mathematic proof of their correctness and efficiency, and experimental measurements that confirm the theoretical results. In the case of P2P, the experimental study in real environment is most times practically impossible, because of the enormous volume of data and involved resources that are required. For this reason simulation environments are employed in order to approach best real life conditions.

In this work, we present a framework that is able to efficiently and scalably support millions of physical cooperating nodes running different protocols and diverse personalized datasets on different levels of granularity. We present the extension of D-P2P-Sim, the so-called D-P2P-Sim$^+$ framework that is able to deliver and to provide two key features:

First, D-P2P-Sim$^+$ supports and facilitates the deployment of very large-scale (multiple million) P2P node systems. It also includes simple, easy and organized tools to achieve the collection of numerous statistics and execution of web-scale concurrent queries, i.e. implementation of node failure and node departure recovery strategies for best simulation of real life conditions.

Second, its modular design enables the use of multiple indexing/processing engines, allowing different applications or even groups of users within the same application to customize their processing needs. All these new features are integrated in the same user friendly environment that provides extensive statistical results. The D-P2P-Sim$^+$ system can simulate 600,000+ nodes in a stand-alone environment and millions of nodes as it can be executed in a distributed mode within a researcher's laboratory. In this sense, it takes advantage of the available resources from multiple computers that usually exist in a typical laboratory or a research network.

D-P2P-Sim+ brings forward appropriate tools and ready to use infrastructure in order to facilitate the researchers' tasks in-line with the framework. Multiple protocols have been simulated in order to show that it takes couple of hours to initialize a network of 100000 nodes at a middle range configured PC. Moreover makes available a framework so that researchers can develop, share and re-utilize failure detection and handling scenarios during their research on P2P network protocols.

Distributed simulations with D-P2P-Si*m*+ can be designed and delivered with the same framework in different environments but with the same single setup. We show that only a small scale laboratory environment (5 computing machines) is an adequate environment to deliver multi - million node simulations at ease. Furthermore, we present the roadmap and demonstrate the results to deliver a small scale laboratory setup to a successfully verified experiment of six (6) protocol evaluations at the PlanetLab platform (http://www.planet-lab.org) with limited effort given resources available. PlanetLab constitutes a world wide research network that supports the growth and implementation of new services for networks.

The rest of the paper is as follows. Section 2 discusses related work and motivation. Section 3 presents the architectural details in depth. Next, section 4 discusses failure scenarios and recovery strategies on large number of nodes networks. Section 5 discusses experimental results showing examples for P2P networks with up to 2 Million nodes using a variety of lookup protocols. Section 6 shows that D-P2P-Sim+ is easily executed in the PlanetLab environment without need to change the protocol. Finally, section 7 concludes the paper.

## Motivation and Related Work

The following subsections describe the related work as well as the reasons motivating us to design a novel distributed framework.

## Handling IoT large amounts of Data

RFIDs technology is being introduced in Supply Chain Management, for constant monitoring of goods, real-time sales information collection and processing. As a result huge number of data is being generated. Building a system to store and process this information adequately is a challenging technically and financially.

As an example suppose that an international company that transports goods tagged with RFIDs is storing all information received by all tags product position, movement, description etc. Every product and its movement is an entry to the system. The company is transporting millions of goods through several locations and the information and its history needs to be stored. In such cases the administration might be interested to know for a specific product the number and places of delay for all product transportation within the last years. In such case each product is a tuple (ProductID, location, time, expected_time (derived attribute), description). Each tuple might reside in different node, based for example on the location of the RFID reader that collected it. Based on the information, location of product delays can be identified and further examined:

*SELECT ProductID, location*

*FROM Nodes*

*WHERE time>expected_time*

As another example we assume that the company stores information regarding all customers that are using their mobile phone to buy products worldwide. Each customer is a node that maintains information as a tuple (UID, number_of_items_purchased, time_of_purchase, location_of_purchase, profile_user). Tuples are stored based on the place of purchase. As customers are purchasing goods information is inserted to the system. The company would be interested to find answers to queries like:

*SELECT COUNT(UID)*

*FROM Nodes*

*WHERE number_of_items_purchased >3*

*AND time_of_purchase in* $[t_1,t_2]$ *AND product=productID*

Not only querying the information is crucial for the participants of a supply chain but also taking the results on time. Major decisions i.e. about purchasing, transporting and producing goods are based on the information retrieved. Moreover, usually participants in different stages of the supply chain as well as customers have signed service level agreements in order to guarantee the timely delivery of goods and their quality. The violation of these quality agreements result in a penalty to compensate. The timely extraction of useful information from the data is hence of paramount importance.

That which would seem a trivial query for a traditional database, is not so, when enormous sets of events/data have been produced from distributed devises. Usually in these environment a new approach in storing and processing the data is adopted: P2P systems as a storage solution provides low cost in sharing of the supply chain information while each node participating in this sharing can be independent, autonomous and use privacy preserving mechanisms to define visibility of data. Range queries performance in a P2P network can vary therefore several overlay protocols have been proposed that improve performance based on different factors imposed by applications and data. The selection and/or creation of the appropriate node based protocol is crucial because it directly influences the response time of the queries hence the satisfaction of the SLAs. Experiments need to be done to decide the appropriate P2P solution settings, and such experiments are not feasible in real environments because of the enormous data collected and the required resources.

Hence, in such IoT applications, where we collect information from several resources that are distributed and each one has an independent set of information, we need tools that can facilitate the development of distributed systems capable of handling queries addressing multi-million nodes. The most prominent way to make decisions upon building such systems is the use of simulators.

In this paper we propose a simulation environment that contributes in taking decisions regarding the choice of the protocol in storage P2P systems. Our contribution is that besides scalability our simulator provides a systematic mechanism to gather statistics. Moreover it simulates a real life P2P storage system realistically taking into consideration nodes departure and failures thus producing accurate results.

## IoT and Personalization

IoT and Web 2.0 technologies (Brusilovsky, Kobsa, Nejdl, 2007; Brusilovsky, & Maybury, 2002) make it possible to deliver personalized views or versions of web data, improving usability and thus productivity.

Adaptation features enable personalization of the interface and the flow of business processes to the different characteristics and role of each internal user. The user profile records information concerning the user and his knowledge state based on the location, the device, the preferences and other information collected by different sensors. User profiles can determine personalized views and they can be distributedly stored across several servers.

This information is vital for the system's operation according to the user's needs and preferences. However, management, recording and manipulation of user profiles at large scale Web Systems can be tedious as multiple replication of Web Systems does not coherently support synchronization of user activity (Wang, Sharman, & Ramesh, (2008). Users transparently enter the Web System at different servers and, as a consequence, their profile can be distributedly stored across several servers.

The explosion in adoption of Web data modeling (e.g., XML and XML Schema (Bertino, Guerrini, & Mesiti, 2004)) has brought forward critical issues in the process of knowledge personalization in Web Services. In this context, service personalization is another issue to be faced, with fine-grained user management being strongly required.

The process of combination for fine-grained multidimensional user models with knowledge representation and management techniques for making Web Services knowledge-aware usually involves a number of Web Services distributed in a network of numerous nodes (Cuzzocrea, 2006). Thus, a distributed approach would be highly useful to efficiently process the amount of profile data.

## D-P2P-Sim+ vs. Previous Solutions

Contemporary distributed systems involve very large populations of commodity nodes independent of the specific underlying architecture: BitTorrent has reported stats where over 20 million daily active users download over 400,000 torrents on average and the number of users is already 100+ millions. Other P2P-based networks like Live Messenger report 300+ millions of monthly active users online (Windows Live Messenger: A short history, 2010) and 25+ million nodes (peak online time) in the Skype chat network (A. S. report on Skype usage, 2011). On the other hand, the processing and retrieval needs per application range vastly: Social and web advertising sites perform map-reduce based intensive processing to identify trends and mine useful information (e.g., (Shmueli-Scheuer, Roitman, Carmel, Mass, & Konopnicki, 2010)), while real-time applications or simple analytics tools require efficient point-range-aggregate queries over bulk personalized datasets (e.g., (Chen, Cui, & Lu, 2011)). Thus, it is also imperative that, besides the required scalability, the platform should be modular enough to provide support for different types of storage/processing engines.

P2P simulators survey (Naicken, Livingston, Basu, Rodhetbhai, Wakeman, & Chalmers, 2007) shows that the majority of the cases studied involved a number of different custom simulation tools. The survey finds that custom made simulators (developed per protocol or in some case utilized by a particular research group) far outnumbers the use of known ones. Resulting, comparison and replication of evaluation procedures and outcomes is very difficult. Additionally, developing multiple times the same protocols from the beginning is needed. A key point discussed in the survey is that `the poor state of existing P2P simulators is the reason that much published research makes use of custom built simulators'. The main drawbacks of the simulation mechanisms found were i) no systematic mechanism to gather statistics of the simulation executions, and ii) absence of graphical user interface support to support user friendliness. To cope with the above problem, the P2P simulator (Sioutas, Papaloukopoulos, Sakkopoulos, Tsichlas, & Manolopoulos, 2009) has been presented.

D-P2P-Sim includes extensive and effective mechanism for statistics as well as an effective and easy to use GUI. In this work we extend D-P2P-Sim.

The analysis of Naicken et al. (2007) shows that P2P simulators do not usually support the creation of more than 10,000 nodes (Jagadish, Ooi, Tan, Vu, & Zhang, 2006). Only a limited number of cases conducts experiments with more nodes (Cowie, Liu, Liu, Nicol, & Ogielski, 1999), however even then no more than a few 100,000 P2P nodes, mainly because of memory requirements. Even though there is evidence that since 2007 a greater proportion of research is being tested with real systems and real trace data (Basu et al, 2013), the scalability of P2P simulators has not been shown in wide scale. Recently, an extension of Peersim (Montresor & Jelasity, 2009) has been presented (Mayer et al, 2012), showing that it is possible to schedule parallel executions of Peersim in order to take advantage of multiple cores within a processor. However, the limitations of the maximum number of nodes remains in each Peersim execution instance. Moreover, no results on the scalability of the proposal (Mayer et al, 2012) have been shown yet to the authors' best knowledge. SimGrid (Casanova et al, 2008, Quinson et al, 2012, & Bobelin et al, 2012) has been reported with results of 2Million nodes on a 12-core processor; this however brings the simulated number of nodes below of 200,000 P2P nodes per processor core. In the SSNG simulator (Teng et al, 2013) the maximal number of (super) peers achieved has been up to 1 Million using the available hardware. The presented framework, using lightweight and efficient collection mechanisms can easily reach >600,000 P2P nodes in a typical Pentium 4 stand-alone PC and multi-million P2P node in a small network that every research laboratory has.

P2P simulators fail to include real-time registration of multiple measures as well as inline features to simulate easily and seamlessly (a) randomized node departures and (b) node failures (and node failure scenarios) to depict more realistically real life conditions (Naicken et al. 2007; Cowie et al. 1999). A work on complex queries has been presented using a hybrid overlay that integrates a structured P2P system with an unstructured one (Lai & Yu, 2012). Researchers usually customize simulators and deliver separate development to facilitate such experimental evaluation (Jagadish et al. 2006).

## Basic Architecture

In this section, the basic architectural features are presented in depth. The figures following depict the main parts of the framework environment, how the packages are related and the design decisions that have been made. The modular architecture allows the researcher to develop quickly and using a clear approach every new protocol that one might want to test on large scale simulations. The simulator is organized in packages as shown in Figure 1.

Four main modules can be distinguished: The message passing environment, the overlay network and protocol instance that is implemented each time; remote services in order to enable distributed execution and simulation; statistical tools and administrative management. The code of D-P2P-Sim+ framework is organized in packages in order to facilitate development.

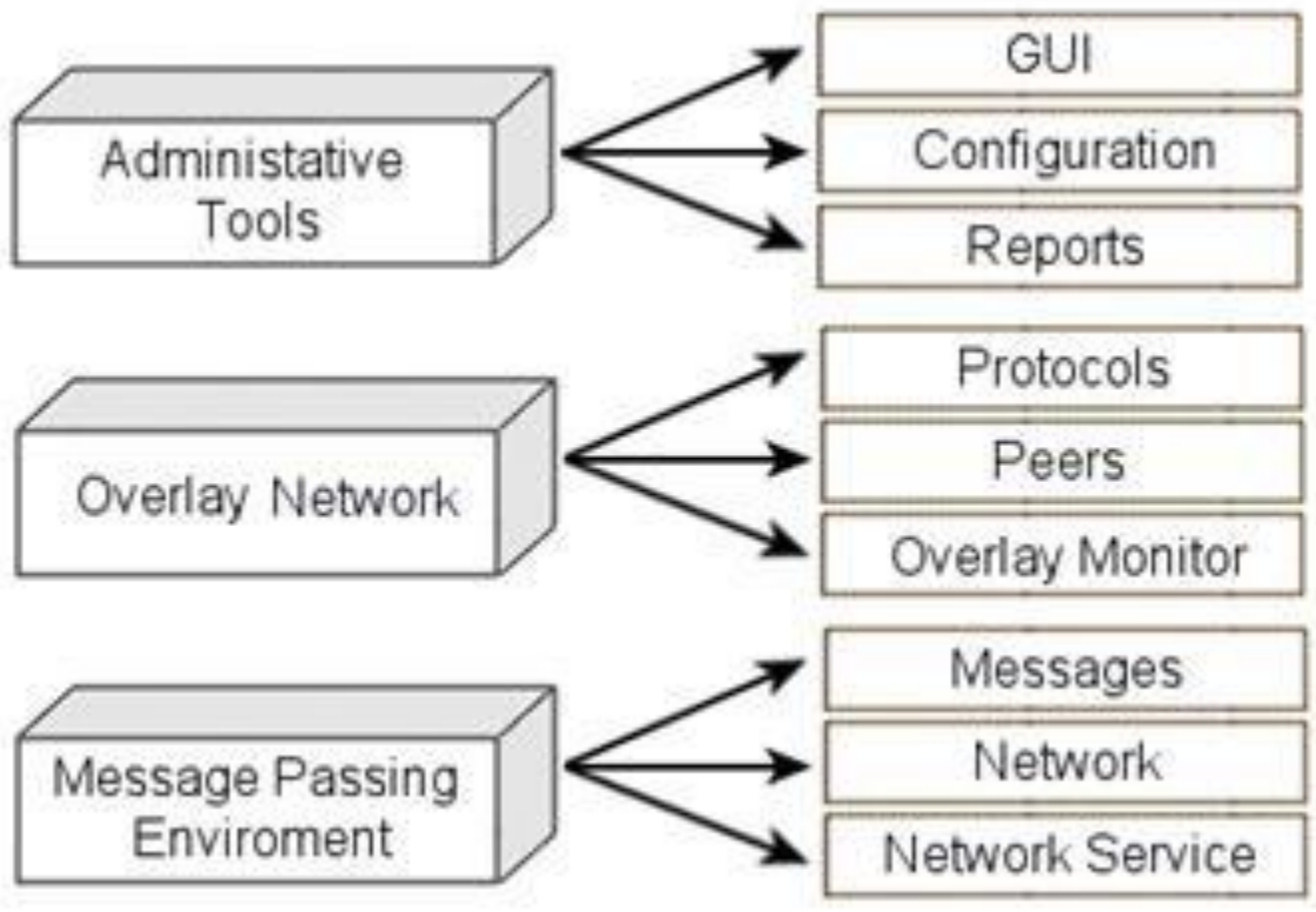

*Figure 1. Simulator main Packages*

All these new features are integrated in a user friendly environment that provides extensive statistical results following the paradigm of (Sioutas et al. 2009). Details on failure recovery strategies, statistics and GUI will be given in the following.

## Message Passing Environment: Configuration Aspects and Network Services

The basic architecture of the Java P2P simulator (see Figure 2) is based on the message passing environment. The environment is solidly based on peers exchanging messages in order to build the overlay network and to carry out the search, insert and delete operations in the P2P network.

Our implementation strategy has determined Message and Data separation of concerns and as a result respective classes have been designed and developed. Any Message object holds the network device/sensor id either serving as a transmitter or receiver. Additionally, the operation that the message serves is also kept within the object. In our solution any Message is coupled to a Data object that includes all its valuable information.

Furthermore, the network's behavior is simulated through dedicated Network class objects. The idea of this object is to deliver the streaming behavior of networking formulating, maintaining and orchestrating buffers i.e. a type of ad-hoc queues of Messages. Any Network object is able to send and receive Messages of different flavors i.e. broadcasting, messages to a specific Node or group of Nodes to serve all possible needs during simulation setups. As it is expected within a Network object all different types of messages “travel” i.e. "search", "insert" or "delete" operation messages. The simulator proposed, for any given message in the network, is able to keep a log record keeping track of its type, sender and recipient to support further statistical analysis if needed. The produced log messages are used by the graphical user interface to show how an operation is incrementally completed.

The network is responsible to notify a node of an incoming message and also is able to reply to queries about messages in awaiting lists (the buffers already mentioned above). The broadcasting message may be used during construction steps of the overlay under simulation in order to send initialization messages to ranges of peers if needed. Broadcast is not used during any search operation in order to avoid flooding consequences.

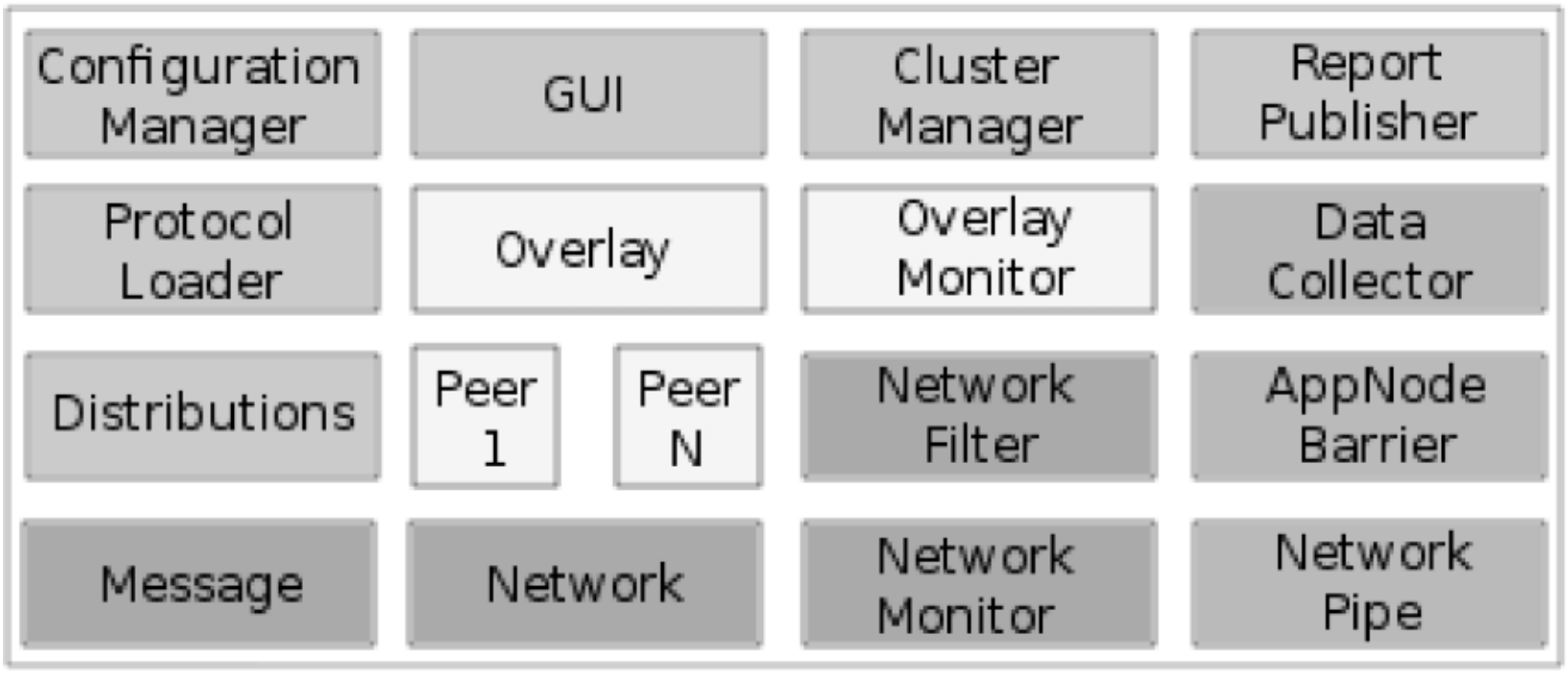

*Figure 2. Message Passing Environment: Peers, Overlay and Network layers.*

## Admin Tools

To support a GUI for multiple protocol implementations and to allow wide customization in testing scenarios and collection of metrics, a number of *administration tools* have been packaged into the simulation framework. *GUI* allows the high level researcher and protocol designer to perform protocol testing without involving source code at all. Furthermore GUI facilitates validation of protocols by independent researchers in an easy and straight forward UI functionality that incorporates collection and presentation of metrics. Admin tools have specifically been designed to support reports on a collection of wide variety of metrics including, protocol operation metrics, network balancing metrics, and even server metrics. Such metrics include frequency, maximum, minimum and average of: number of hops for all basic operations (lookup-insertion-deletion path length), number of messages per node peer (hotpoint-bottleneck detection), routing table length (routing size per node-peer) and additionally detection of network isolation (graph separation). All metrics can be tested using a number of different distributions (e.g. normal, weibull, beta, uniform etc).

```
< distribution>
  < random>
    < seed>1</seed >
< /random>
< beta>
   < alpha>2.0</alpha >
    < beta>4.0</beta >
< /beta>
< powerLaw>
   < alpha>0.5</alpha >
   < beta>1.0</beta >
< /powerLaw>
< /distribution>
```

*Figure 3: Snippet from config.xml with the pre-defined distribution's parameters setup*

Additionally, at a system level memory can also be managed in order to execute at low or larger volumes and furthermore execution time can also be logged. The framework is open for the protocol designer to introduce additional metrics if needed. Furthermore, XML rule based *configuration* is supported in order to form a large number of different protocol testing scenarios (see Figure 3). It is possible to configure and schedule at once a single or multiple experimental scenarios with different number of protocol networks (number of nodes) at a single PC or multiple PCs and servers distributedly.

## Overlay Scalability

Apart from the fact that a number of packages facilitate development, a number of different protocols are available as sample code and executables in order to allow familiarization with the development and usage of the simulator: Chord, Baton*, Nested Balanced Distributed Tree (NBDT) (Sioutas, 2008), NBDT* (Sioutas, 2008), R-NBDT* (Sioutas, 2008) with advanced load distribution and ART (Sioutas, Papaloukopoulos, Sakkopoulos, Tsichlas, Manolopoulos, & Triantafillou, 2010). Moreover, the respective abstract classes and programming steps are depicted also at a simplistic *dummy protocol* in order to guide the programmers that first use the simulation framework proposed. Figure 4 depicts main performance measurements.

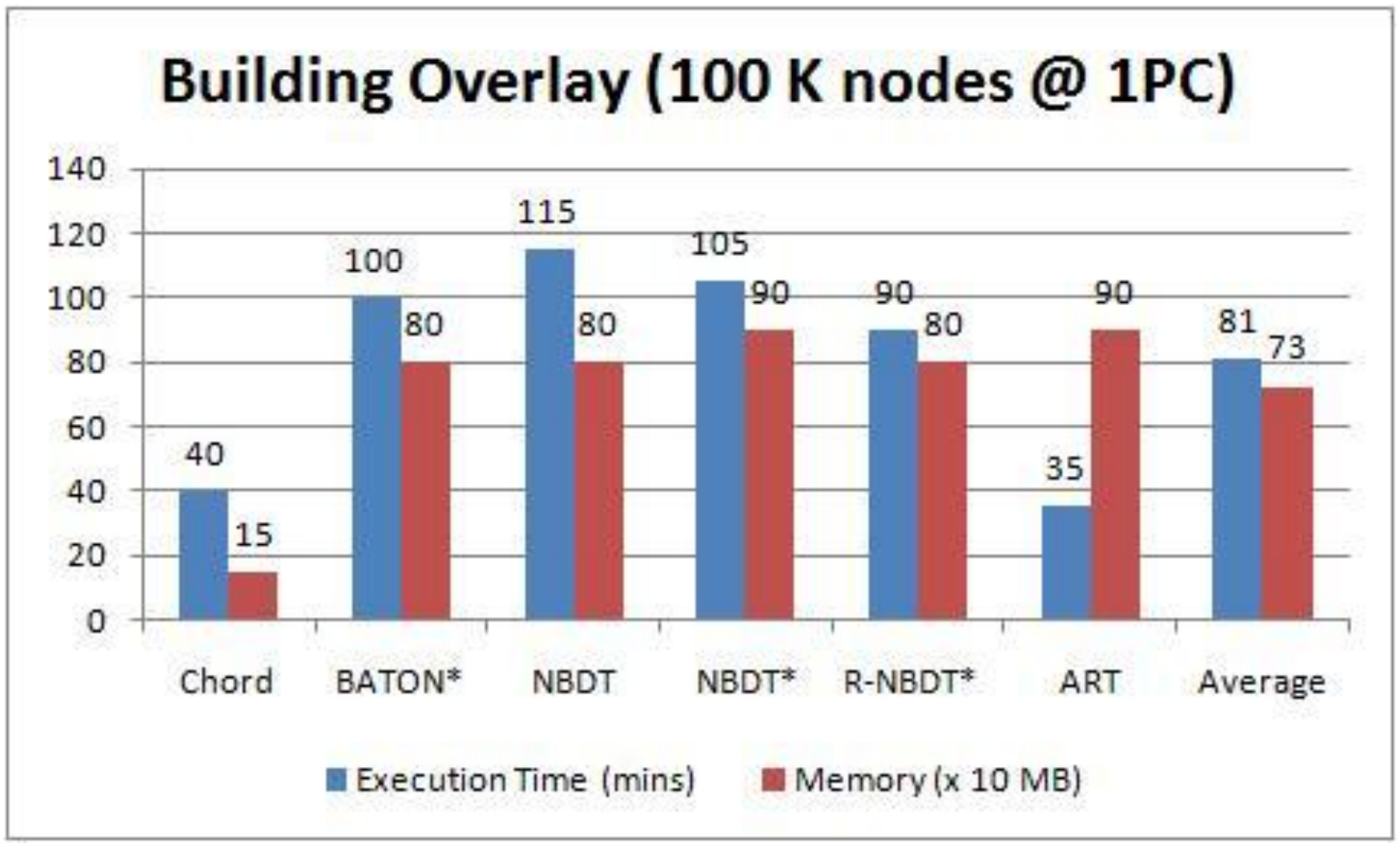

*Figure 4. Protocol Support: Execution time and memory requirements*

The framework is particularly designed to allow large scale experimental evaluation of node based protocols. The incorporation of metrics collection methods into the framework reveals the space and time consumption of each implemented p2p protocol and this fact minimizes both the memory needed (and frees up space for more nodes to be executed) as well as the response time of the p2p Handling method we finally chose. Moreover, distributed scenarios allow multiple computers to participate in the same experiment increasing radically the number of nodes simulated.

## Node Failure and Departure Strategies and Statistics

Node failures and departures are important in order to achieve real life conditions during simulation. The presented framework supports new operations and services so that it provides services for node failure and network recovery and for node departures and substitutions. Simulators up to now tend to limit themselves to support (a) import of nodes for the creation of overlay network and (b) searching, (c) import and deletion of keys in nodes.

For this reason, we present the operations of a new simulator with all the necessary services to implement strategies for

- the self-willing departure of peer nodes (node departure)
- the failure or their sudden retirement for any reason (network failure, application failure etc.)
- monitoring whether the overlay network is parted after successive node failures or departures
- statistics and management in order to monitor all different strategies natively in the simulator. Statistics include: (i) cost monitoring of importing and retirement of nodes, (ii) the number of failed queries due to node failures and (iii) the number of nodes leading to overlay partition

Self-willing departure of nodes could be simulated following two approaches. It is possible to simulate scenarios that departure of nodes is initiated in parallel, so that random nodes in the network depart simultaneously. A second approach is to set nodes departing in a sequential mode, where each next node departs after another's complete leave. Though the simulator supports both modes, we have seen that sequential departures are not realistic and as a result they tend to hide problems that might appear (e.g. simultaneous departure of a node and its backup node). On the other hand, multiple concurrent departures make P2P protocol designers to deal with such cases.

The classes and packages of the framework include all the necessary code parts to facilitate the researcher in order to detect, control departures and execute simulation using them. The message passing environment is designed to be possible to detect during simulations whether the message recipient is online or not (i.e. it does not take for a fact that the nodes are always available as done in rest of simulators).

Additional details and technical specifications on the failure scenarios related operations are given in the following.

It is common place that sudden departure of nodes without notice can bring large scale problems to routing messages within a P2P network. Such sudden simultaneous departures of multiple nodes makes difficult to failure recovery routing strategies find an alternative path to avoid failure nodes. In such cases, sending messages to the same node more than one time is probable. In order to overcome such cases, the simulator infrastructure includes tools to store all intermediate nodes that a message visited in its path for the sake of the simulation study even in cases that a protocol does not need such information. As a result, all paths can be stored and studied after each message is sent.

Moreover, to support departure and node failures for any protocol, node selection strategies take into consideration exception lists for nodes that have failed while running any distribution for experiments. Simulator facilities allow passing failed node lists and/or departed node lists to preprocessing of distributions utilized to retrieve randomized node ids during experimental runs.

D-P2P-Si$m^+$ determines the state of each node based on the state of each thread implementing it (RUNNABLE, BLOCKED, TERMINATED). However, in order to handle self-willing node departures and sudden retirement of nodes, the node has to pass through different states during simulation, independent from the thread state that implements it. These states are fixed in the class named PeerState that was created as a part of network Overlay package, and they are following: (a) WORKING: online and ready, (b) CANDIDATE_SUBSTITUTE: node substitution by existing node or nodes at the network, (c) VOLUNTARILY_LEFT: node self-willing departure, (d) FAILED: abrupt node failure.

Next, it is crucial to monitor for overlay network partitioning after failure of successive nodes, which results in isolation of nodes. This happens when all routing pointers to nodes outside the isolated partition are broken. In other words, the minimum number of routing pointers that should break in order that a group of nodes is separated from the overlay network equals to *S* as follows: S=Σ contacts of all nodes of team - Σ internal contacts between nodes

Furthermore, details on the statistics needed to monitor failures and departures of nodes are: (a) The number of steps (hops) to find the suitable overlay network point for additions in the overlay network. This calculation is realized with the infiltration of message JOIN_RESP from the Network Filter class. (b) Number of steps to find a substitute when a intermediary node wishes to withdraw itself. This calculation is realized with the infiltration of message REPLACEMENT_RESP from the Network Filter class. (c) Number of search queries, additions or deletions of keys that failed, either because the expected node has departed, or because the network has been partitioned and it is not possible to access the requested destination. This calculation is realized with the sum of messages QUERYFAILED_RES from the Network Filter class. In particular, a main target of this framework is to maintain robust collection and analysis of statistical data that results from the experiments at all new failure detection and overcome scenarios strategies. The importance of experimental analysis is of more major importance, because it confirms the theoretical analysis, it elects problems and potential omissions that had not been located during the theoretical study, and constitutes a proof for the proposal that is presented. Moreover, the more realistic the simulation is, the smaller the divergence of experimental results from the real life results.

## GUI Support

Throughout this section the functional specifications and the Graphical User Interface (GUI) of the D-P2P-Si*m*+ framework are described. The GUI (see Figure 5) is organized in such a way that can guide someone easily through the simulation process with no need for any prior knowledge about the simulator.

The GUI includes full management and handling of all the above operations. The GUI gives the user the possibility of setting up simulation experiments for protocols without touching any supplemental files for initial parameters (configuration files) via the appropriate tabs. The tabs of GUI named *Operation* and *Experiments* respectively are the ones that users implement the operations and setup experiments respectively. More analytically, by using the tab *Operation*, users can apply self-willing retirement and failures of nodes according to their wishes, within the limits of the overlay network.

Via the tab *Experiments*, the GUI gives the user the option to execute multiple experiments to test search, addition and deletion of keys, so that it evaluates the protocol of his wish. It is also possible to design experiments for mass self-willing departures or failures of nodes. The self-willing retirement may be set to be done in either successive randomized manner or in batch mode where multiple nodes fail together at the same time.

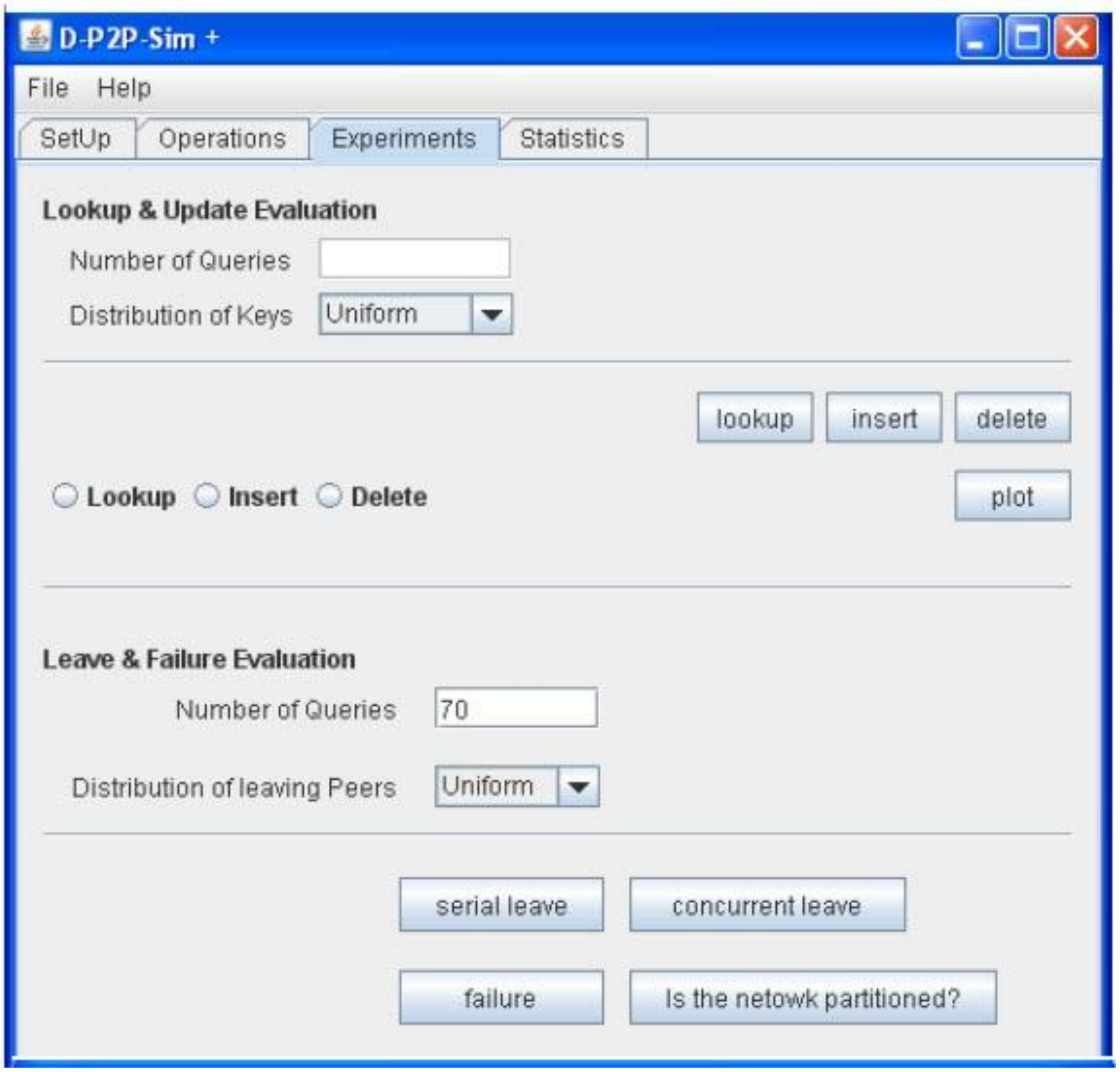

*Figure5. Setup, Operations, Experiments and Statistics Option*

During initiated node failures, one can check if a partitioned group of nodes has been created in the network, using the appropriate button, the so-called *Is the network partitioned*?. Respectively, statistical analysis is available at the *Statistics* tab. Following, further details of the experimental procedures are discussed for different setups.

## Simulation Environment

A standalone experimental evaluation at a typical research laboratory computer can be performed either using the GUI or filling out XML configuration files with the necessary the parameters of execution.

First, we demonstrate the efficiency of our framework to deploy a number of distributed protocols and test their lookup performance. Using a single-PC configuration (Intel Core2 Duo CPU @ 3GHz, 3GB RAM) we simulate 100K node overlays with *six* different protocols: Chord, BATON*, NBDT, NBDT*, R-NBDT* and ART. Results that register the required time for overlay construction and memory requirements are presented in Figure 4. Firstly, we note that even a single typical PC can easily host wide scale experiments. Secondly, our framework easily hosts a wide variety of protocols and manages to very efficiently build a large overlay using minimal resources: At most 1 GB of memory is required for 100K overlay construction and full functionality. Execution times for this mode are also very small, ranging from 35min to at most 115min for NBDT.

The distributed environment (see Figure 6) that was used for the remainder of the experiments consists of a total of 5 PCs (Intel(R) Xeon(R) CPU E5504 @ 2.00GHz, 8BG RAM) running over Ubuntu 9.10. To show the potential of the framework we evaluated the performance of a number of protocols. Exploiting the efficiency and its ability to function in a distributed environment, we executed large-scale simulations, altering the size of the network from 100,000 up to 2,000,000+ nodes with different flavors of protocols.

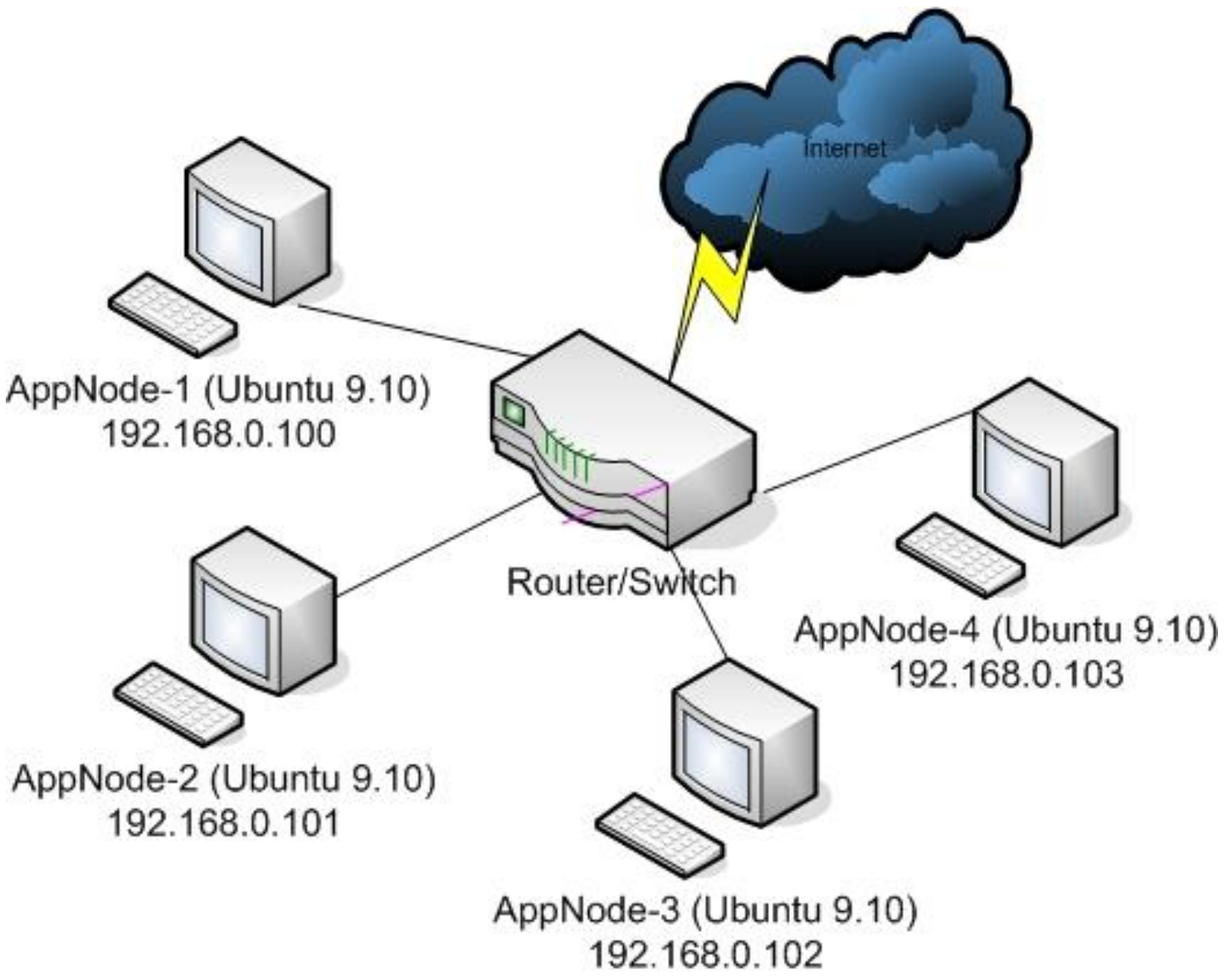

*Figure 6. The Distributed Environment*

Node failure and recovery scenarios have been evaluated as well as node departures and substitutions to test fault tolerance per protocol. For all this spectrum of nodes and fanouts the configurations used, took over 8 hours to deliver full output. Trials have been made with single, and 2-core up to 8-core processors and with 0.5 GB up to 16 GB of RAM memory verified that the simulator scaled without problems and did not need any changes of its configuration. This is also verified for experiments with 1 up to 12 PCs which is a number of PCs expected available at a typical University Lab (WAN experiments are also done on 50 Planetlab points, please see section below). Furthermore the number of nodes simulated scaled near-linearly.

## D-P2P-Si*m*$^{+}$ Experimental Evaluation

Figures 7 and 8 present results for exact-match (or lookup) and range queries respectively while experimenting with two (2) well known protocols: Baton* for up to two million nodes and ART for up to 600K nodes. The performance of P2P protocols is greatly dependent on the average path length between two random network nodes. Figure 7 shows the path used in order to define the exact-match node-positions where insertions and deletion of keys have to be executed. Results verify that query cost for Baton* and ART is logarithmically and sub-logarithmically increasing respectively with the network population. Moreover, experiments show that query costs are decreasing with a fanout increase. In particular, the gain received from the fanout increase is larger when the network becomes more massive. This is expected as the larger the number of nodes, the smaller the tree height becomes with the fanout increase (in terms of rate). As a consequence, we have verified for overlays up to 2M nodes that the cost of search, insert and delete in Baton* and ART protocols is $O(\log_m N)$ and $O(\log_b^2 \log N)$ respectively, where *m* is the fanout factor in Baton*, b is the fanout in ART and *N* is the number of nodes. Note that, according to Baton* p2p structure, original results presented in (Jagadish et al. 2006) are up to 10K nodes.

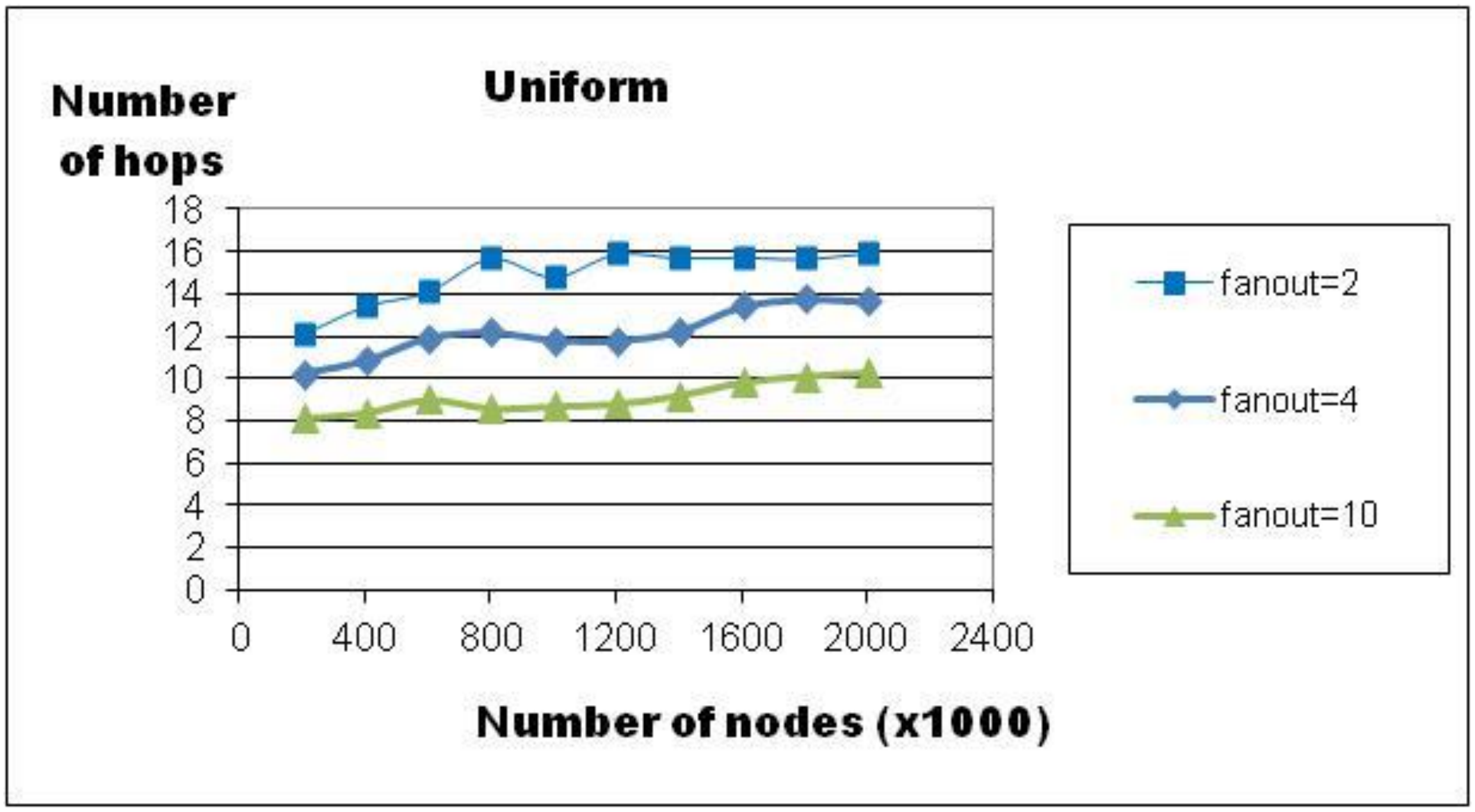

*Figure 7(a). Baton* Simulations with up to 2 Million P2P nodes: Lookup Cost Measure*

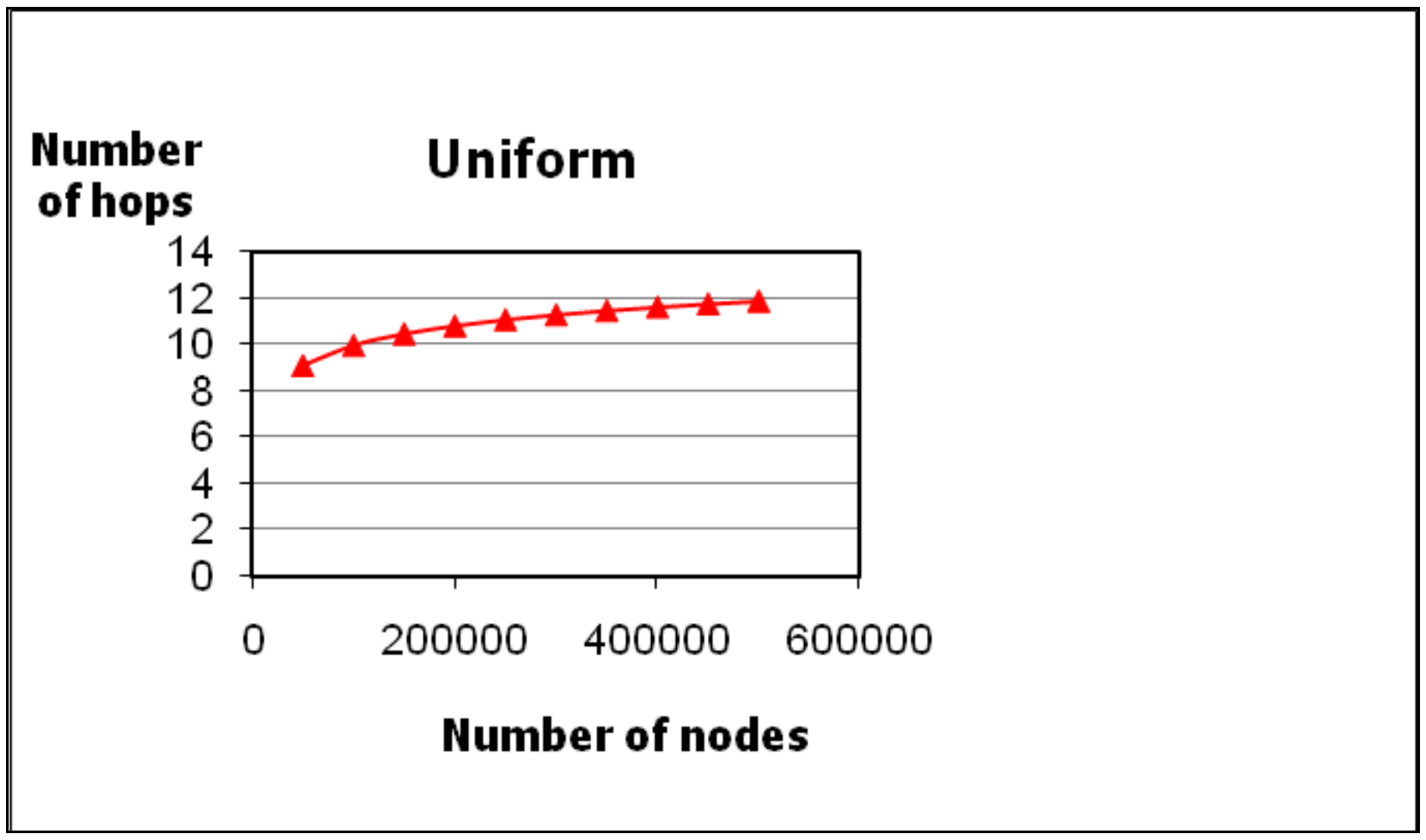

*Figure 7(b). ART Simulations with up to 600K P2P nodes and Uniform Distribution: Lookup Cost Measure*

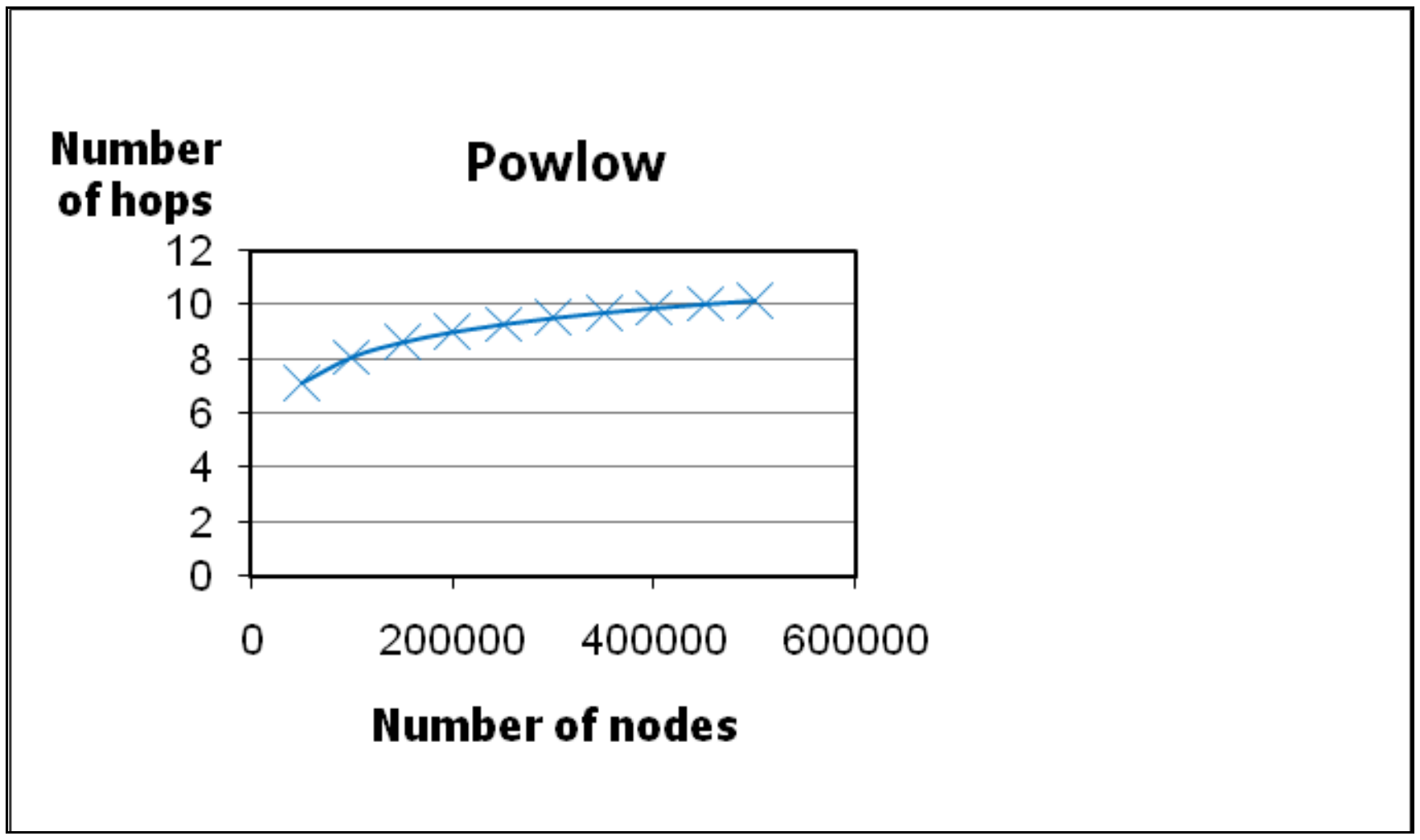

*Figure 7(c). ART Simulations with up to 600K P2P nodes and Powlow Distribution: Lookup Cost Measure*

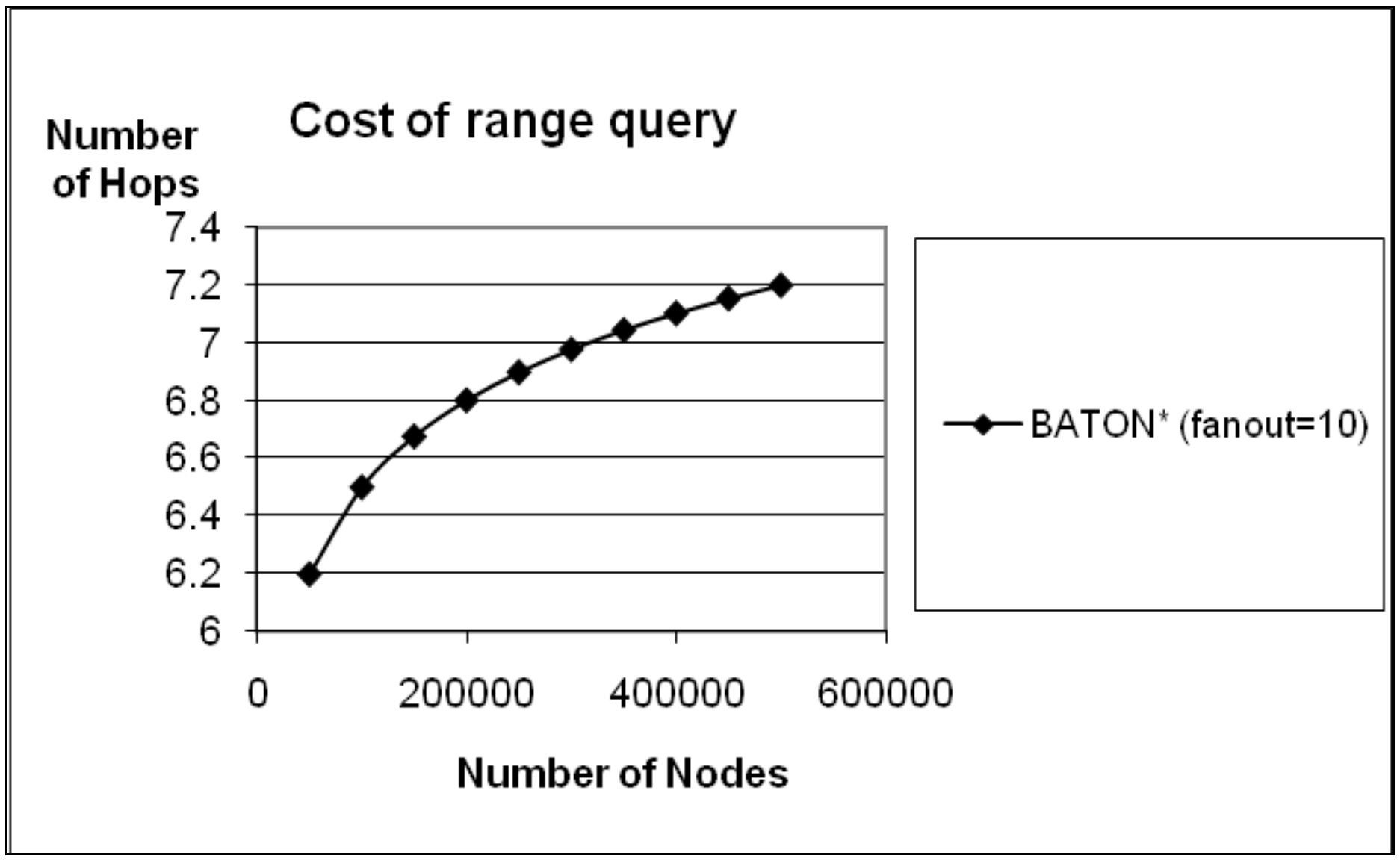

*Figure 8(a). BATON* Simulations with up to 600K P2P nodes: Range Query Avg Cost Measure for arbitrary distribution*

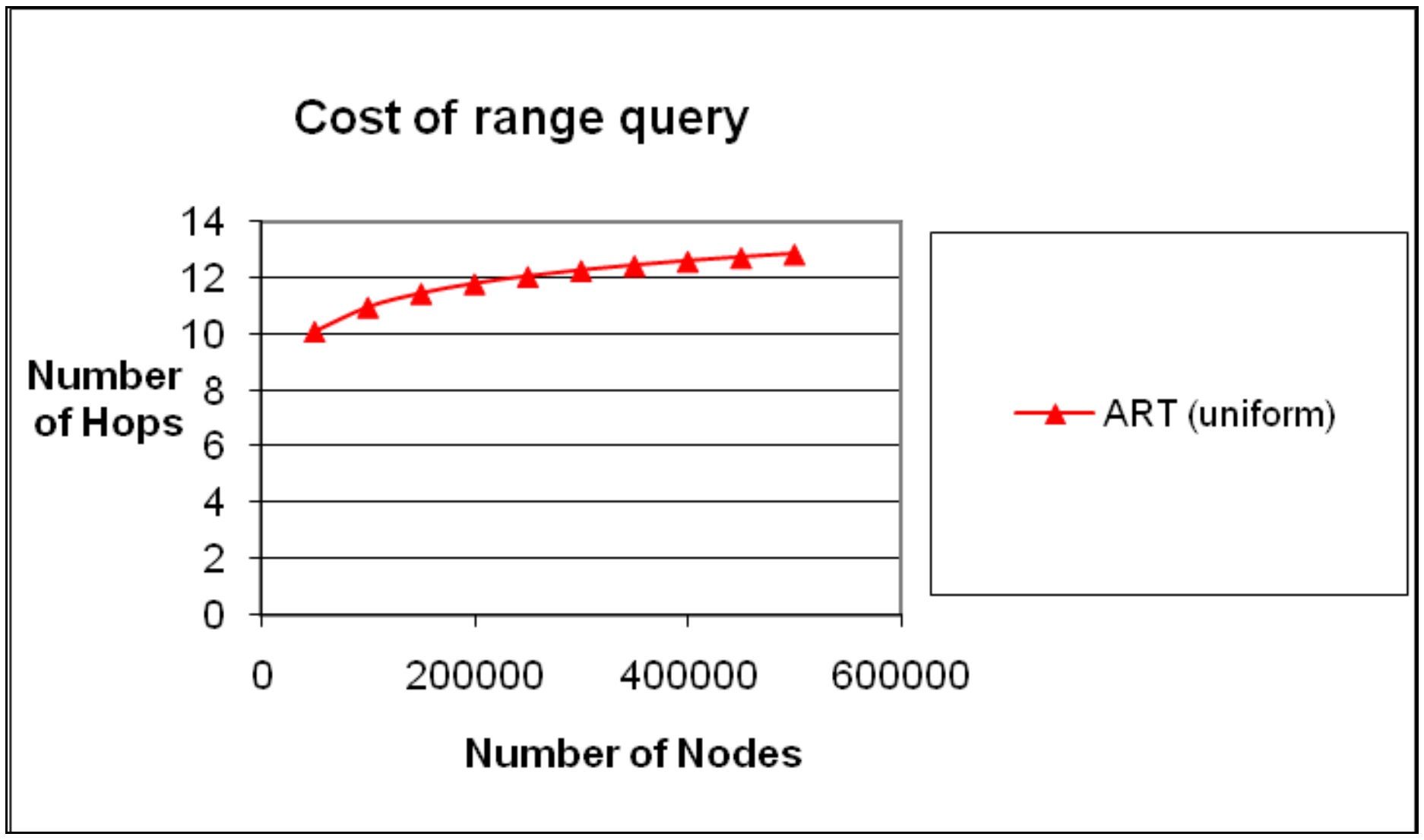

*Figure 8(b). ART Simulations with up to 600K P2P nodes: Range Query Avg Cost Measure for uniform distribution*

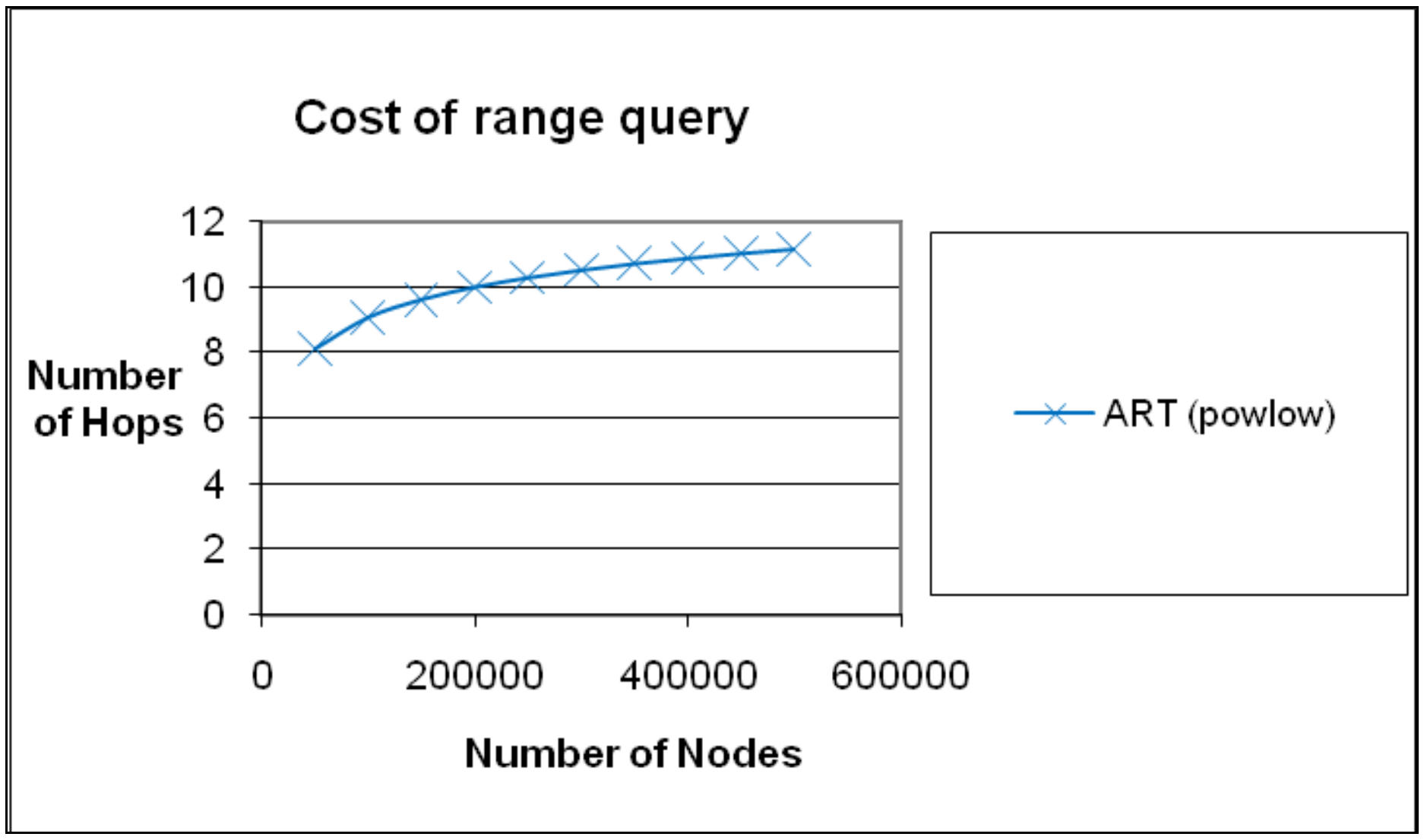

*Figure 8(c). ART Simulations with up to 600K P2P nodes: Range Query Avg Cost Measure for powlow distribution*

Figure 9 shows the average routing table length for Baton*. It is clear that with an increase in the network size, the routing information that has to be maintained per node is also increasing: Increasing the number of nodes, more levels are created in the tree and therefore, the number of neighbors maintained at each node is increasing too. Moreover, a fanout increase leads to an increase of routing table length and, as a result, the cost of updating it. Figures 10 show cost measurements for Updating Routing Tables in Baton* and ART structures for a variety of input distributions. Experiments verify that faster search is possible using larger routing tables

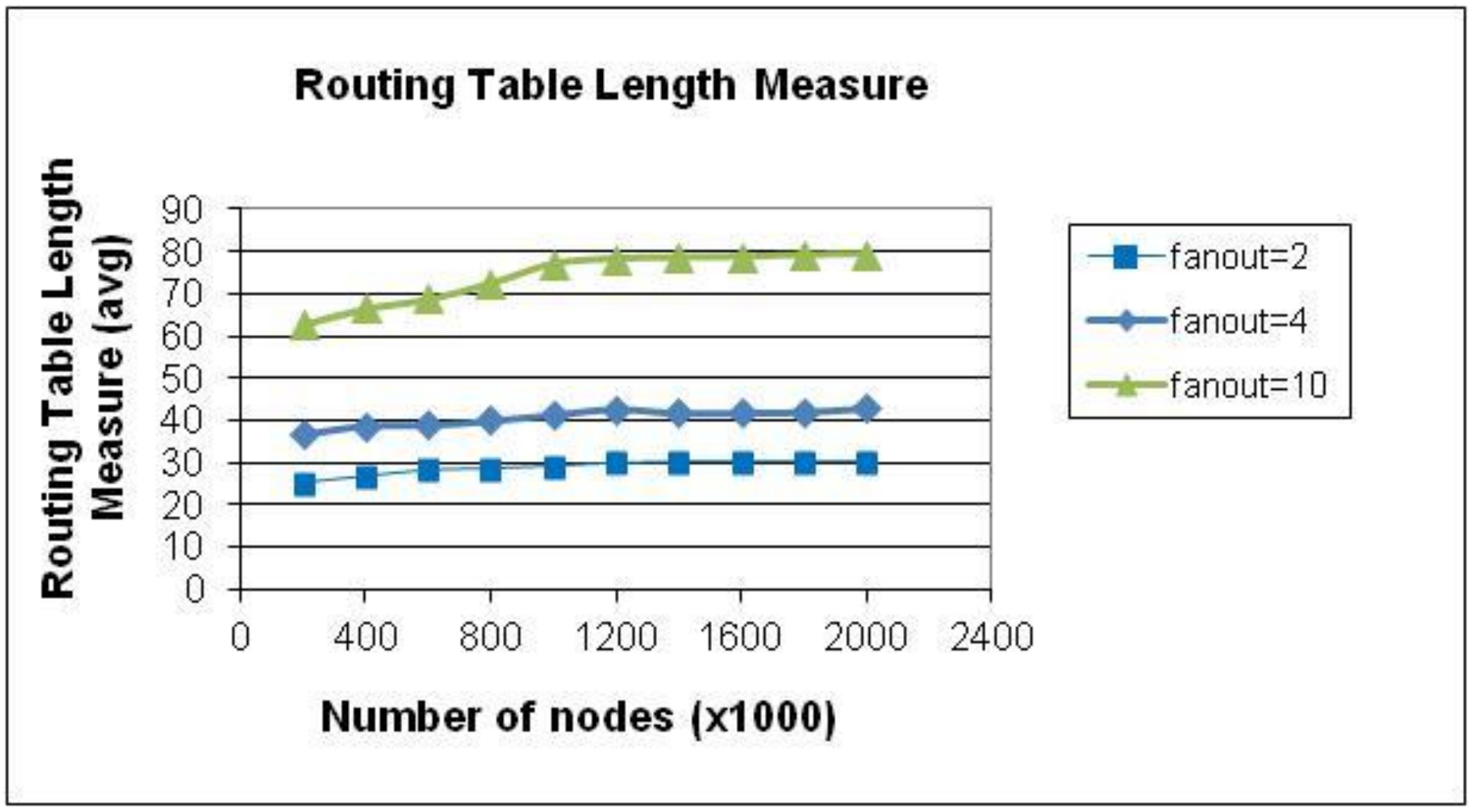

*Figure 9. Baton* Simulations with up to 2 Million P2P nodes: Routing Table Length Measure*

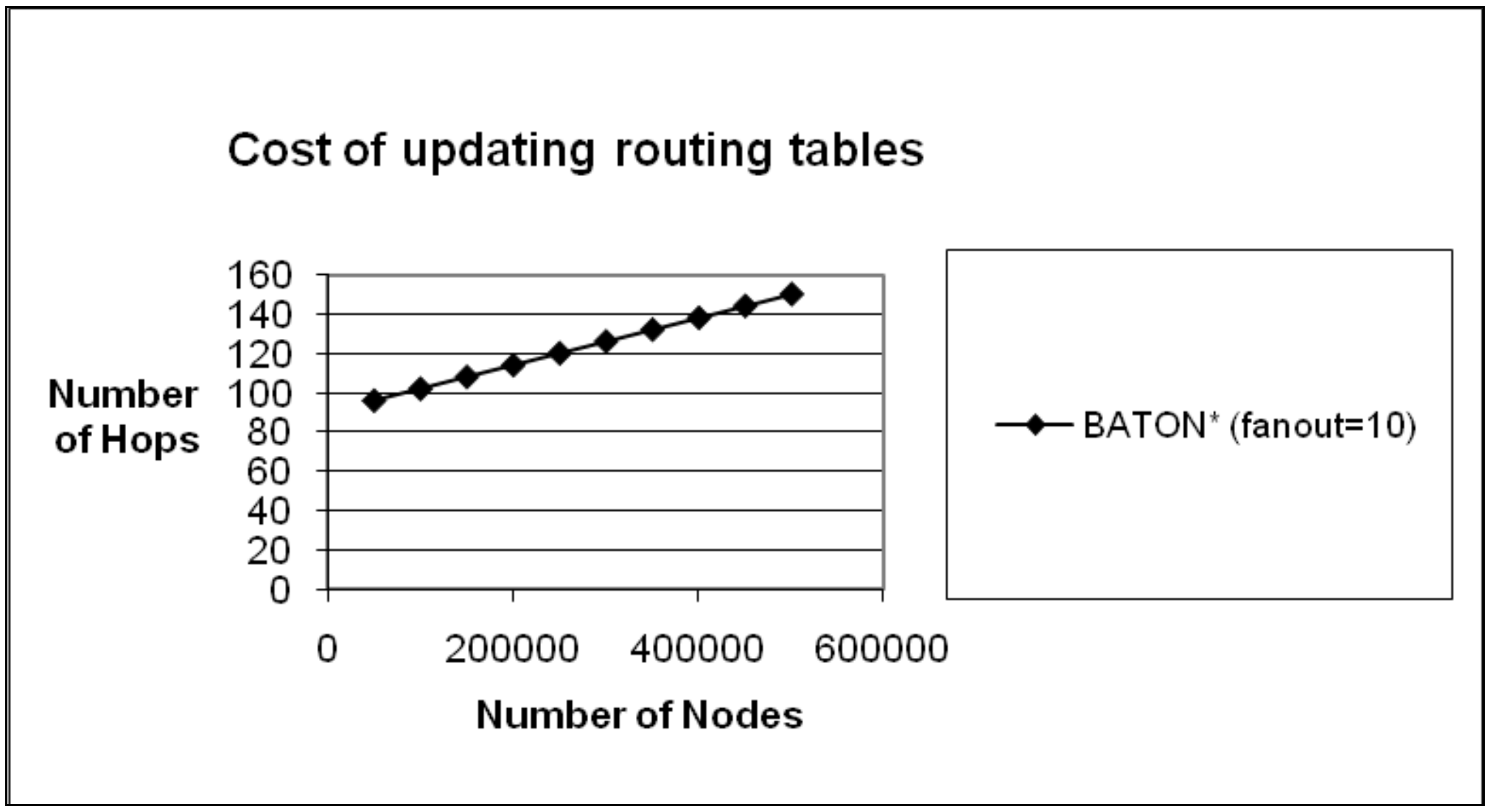

*Figure 10(a). BATON* Simulations with up to 600K P2P nodes: Updating Routing Table Avg Cost Measure for arbitrary distribution*

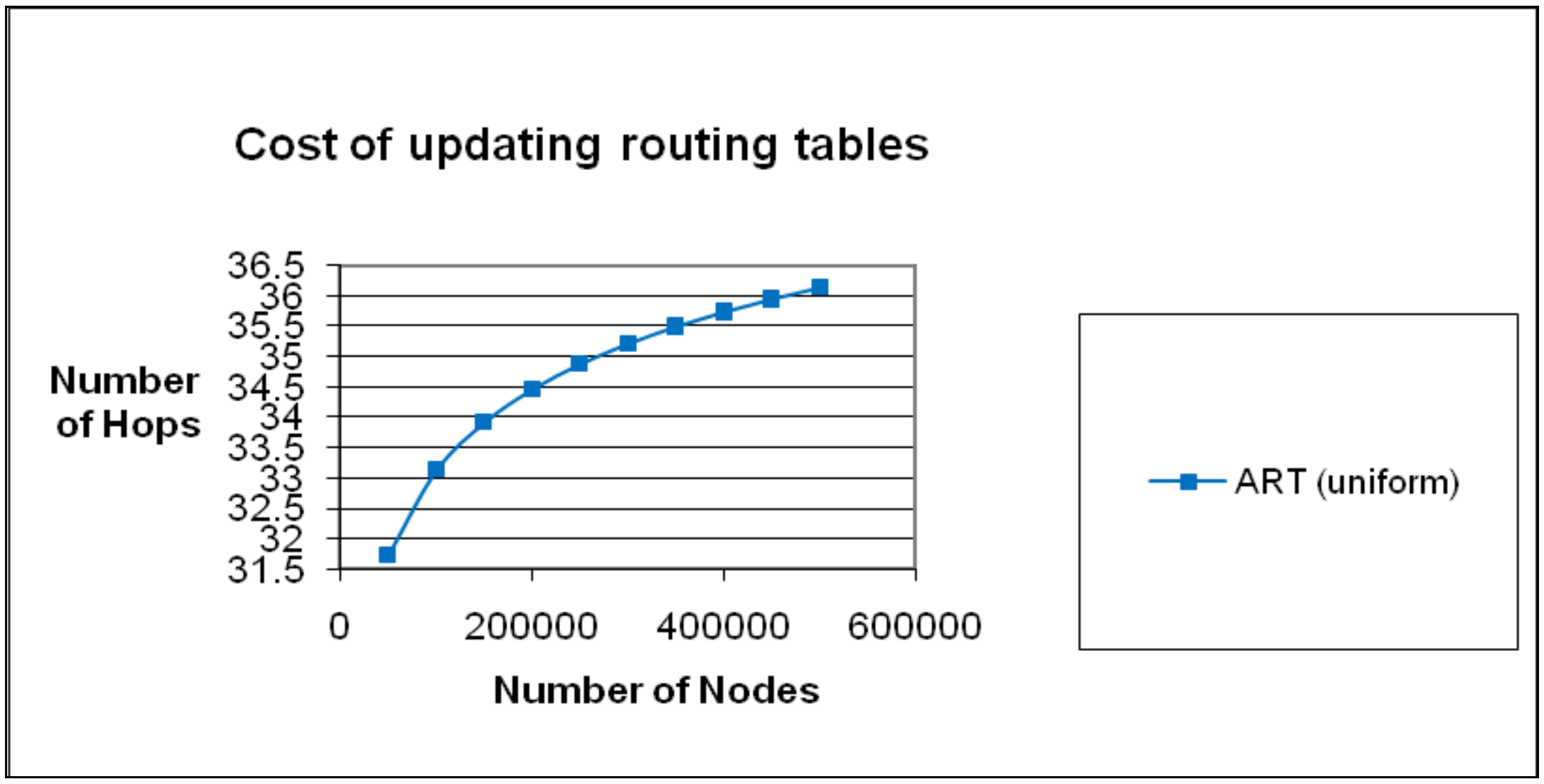

*Figure 10(b). ART Simulations with up to 600K P2P nodes: Updating Routing Table Cost Measure for uniform distribution*

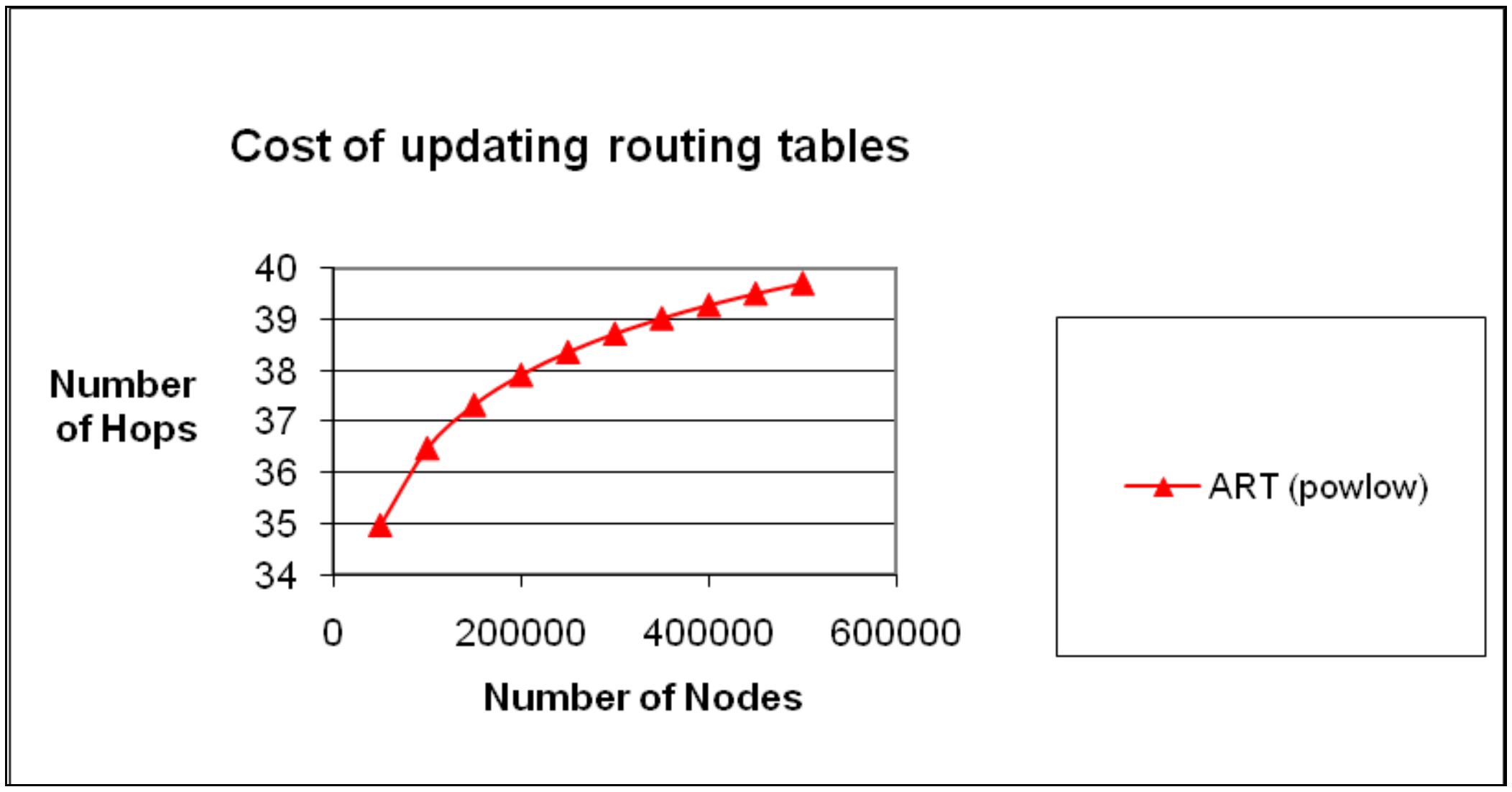

*Figure 10(c). ART Simulations with up to 600K P2P nodes: Updating Routing Table Cost Measure for powlow distribution*

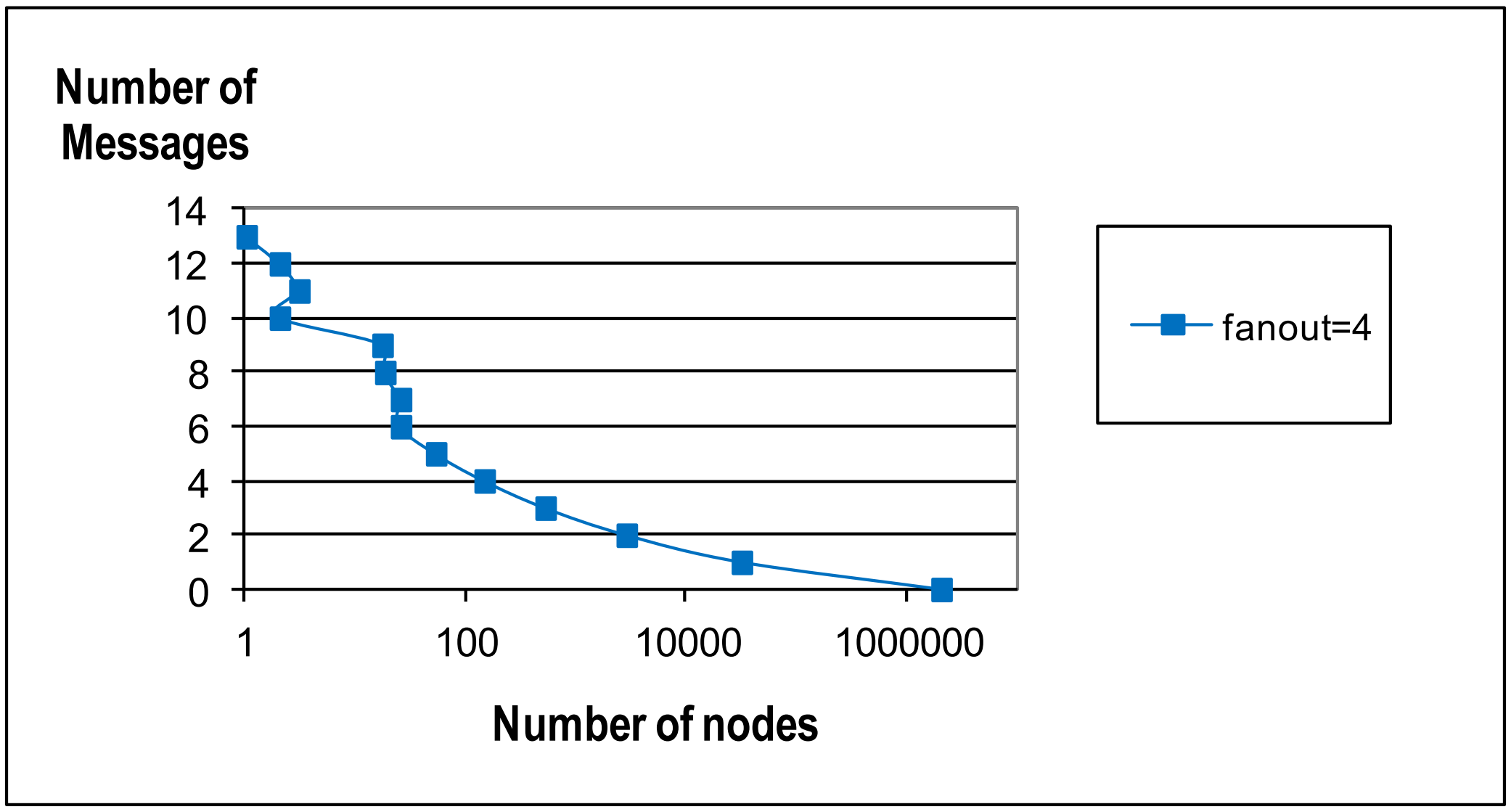

*Figure 11(a). Baton* Simulations with up to 2 Million P2P nodes: Messages per node measure*

Figures 11 present the load that each node receives, counting the number of messages received for any of the operations for node populations up to 2M nodes and for 3K operations. As shown, in both Baton* and ART structures, the maximum number of messages in the experiments reached 13 and this only happens for a single node. On the contrary, 30,590 nodes receive a single message in Baton* structure and ~1M nodes receive 6 messages in ART structure. As a result, we note that both protocols successively balance the load among nodes. This also verifies that D-P2P-Sim+ includes all the necessary features and tools to detect and research on load balancing techniques for P2P network protocols. As discussed in previous section, a list of collectable metrics is available to evaluate and experiment with protocols at application, protocol and system level.

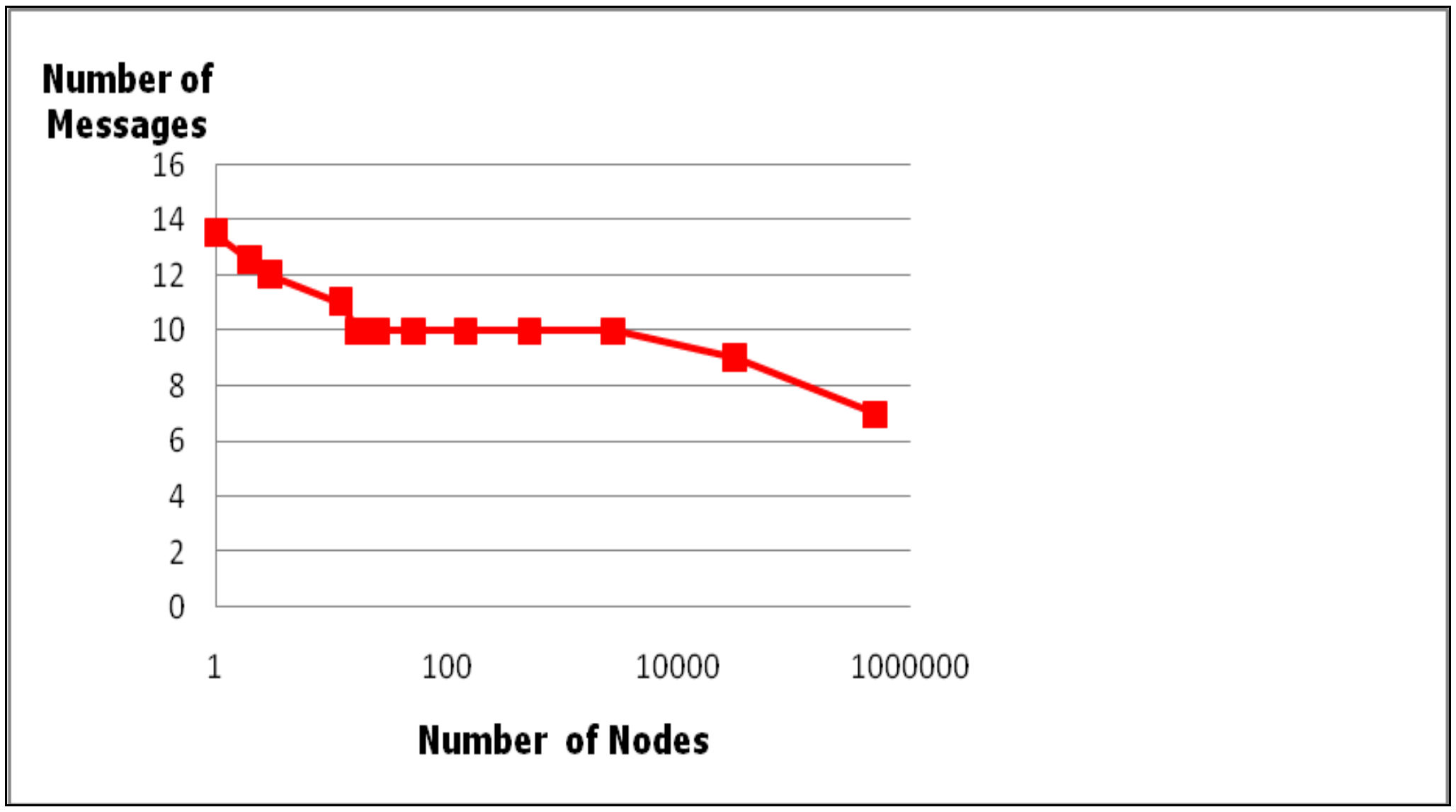

*Figure 11(b). ART Simulations with up to 1Million P2P nodes: Messages per node measure*

We also show that our framework is able to collect connection-specific metrics that can be used to evaluate node-failure and recovery strategies on networks with multi million nodes. The problem that massive failures cause is the invalidation of links among them. As the search procedures have to overcome non-reachable connections, it is hard to choose a path that does not include failed nodes. Queries oscillate inside the overlay until the alternative path is determined. Thus an increase in node failures is expected to result in an increase to search costs. In our setup, a network of 1M nodes is initialized, with a 10% of randomly chosen nodes being assigned to leave abruptly (sudden node death), without rebuilding the network. The experiments continue increasing the percentage of nodes failing in the network by 1% at each round. In each step, the network is checked in order to verify that it is not partitioned into isolated areas that cannot communicate. All the experiments are repeated for fanout 2 up to 10 for Baton*.

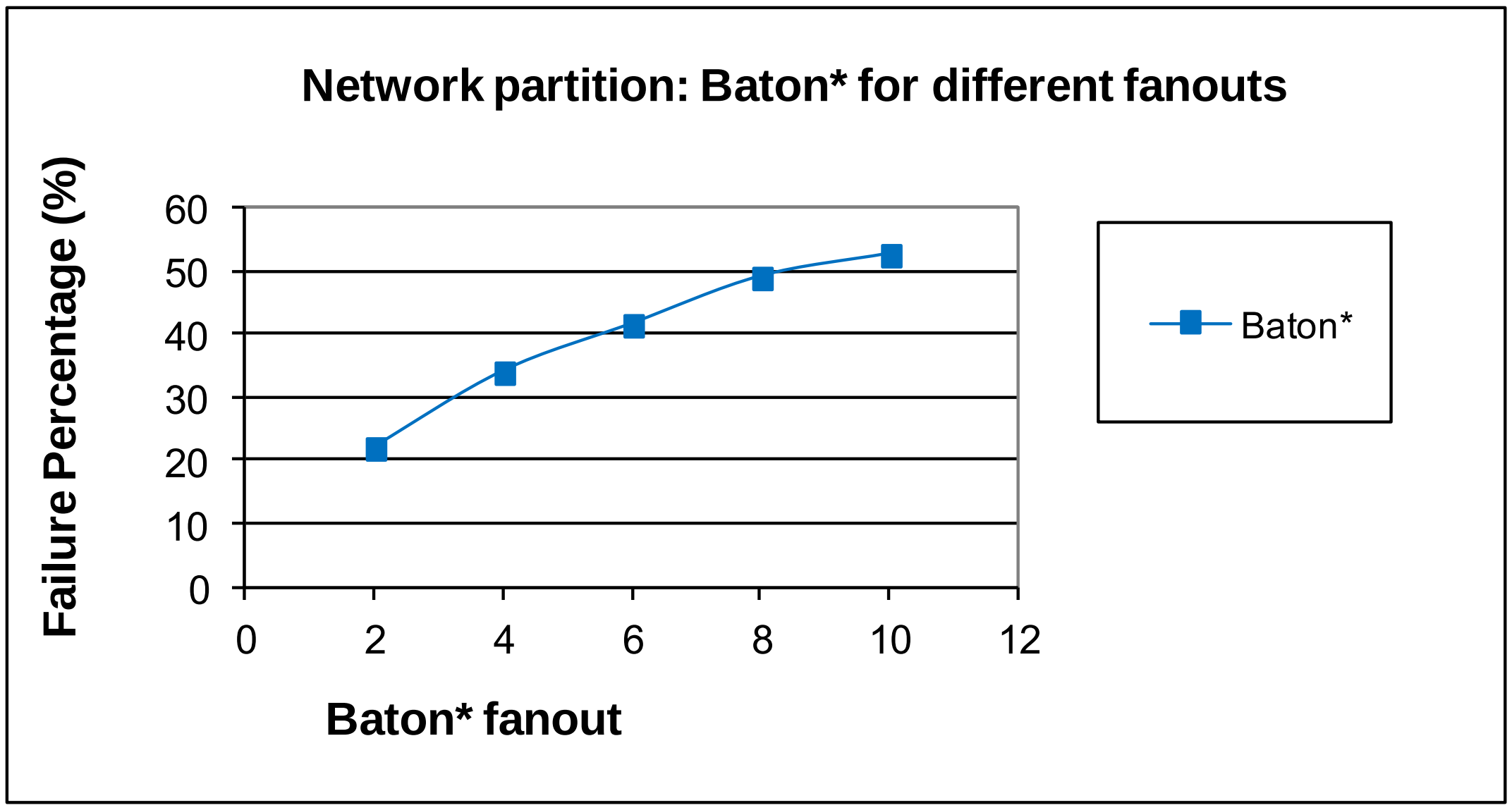

*Figure12. Baton* Simulations with up to 2 Million P2P nodes: failure percentage measure before network partition*

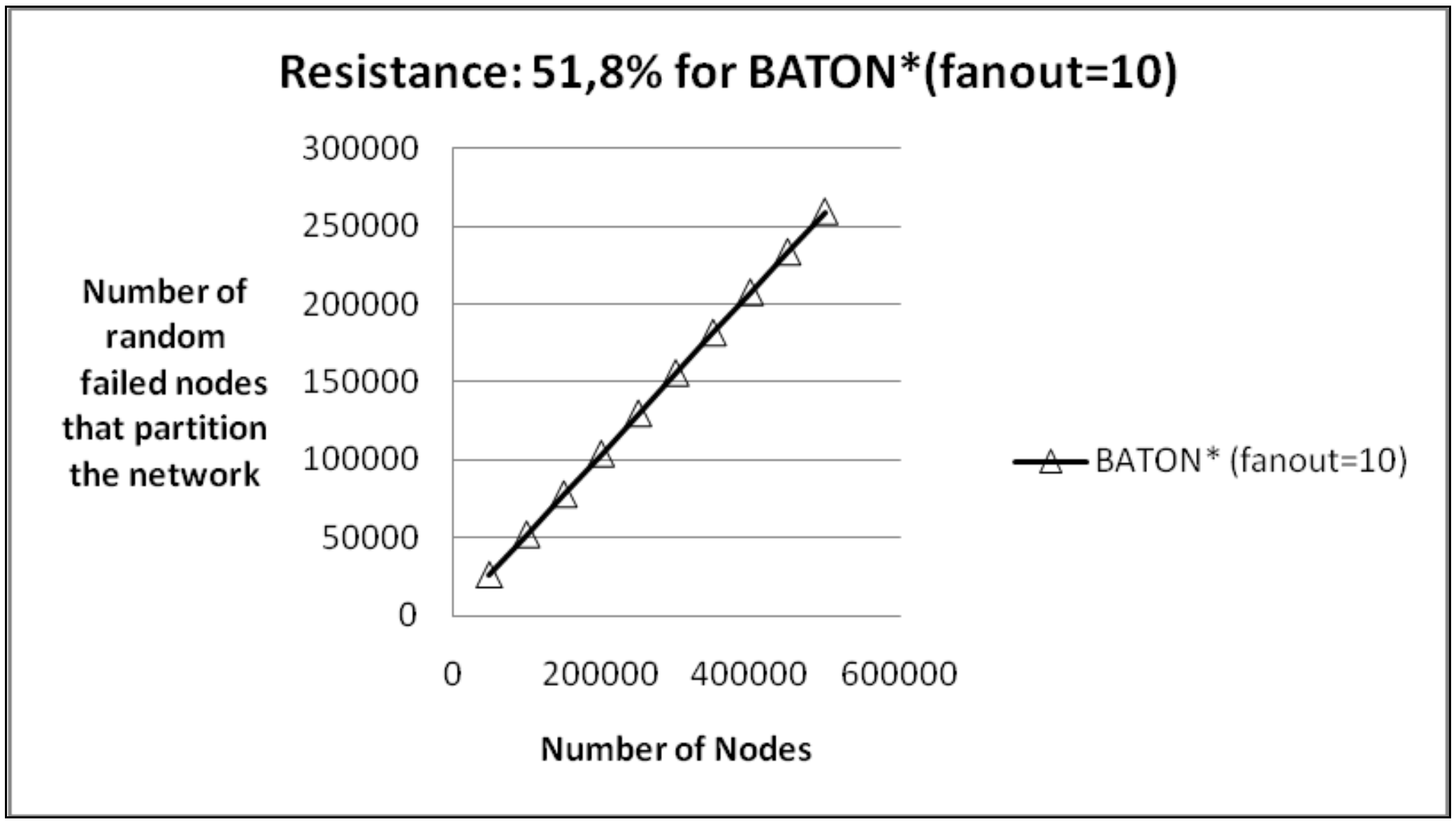

*Figure 13(a). Baton* Simulations with up to 600K P2P nodes: Resistance percentage measure*

Figure 12 shows the average number of nodes that is expected to fail before the network is partitioned. The results verify that the network is resistant to failures when a quarter of the nodes fails for *fanout=2*. A fanout increase results in higher degrees of tolerance. Figures 13, show the resistance percentage measure for Baton* and ART respectively.

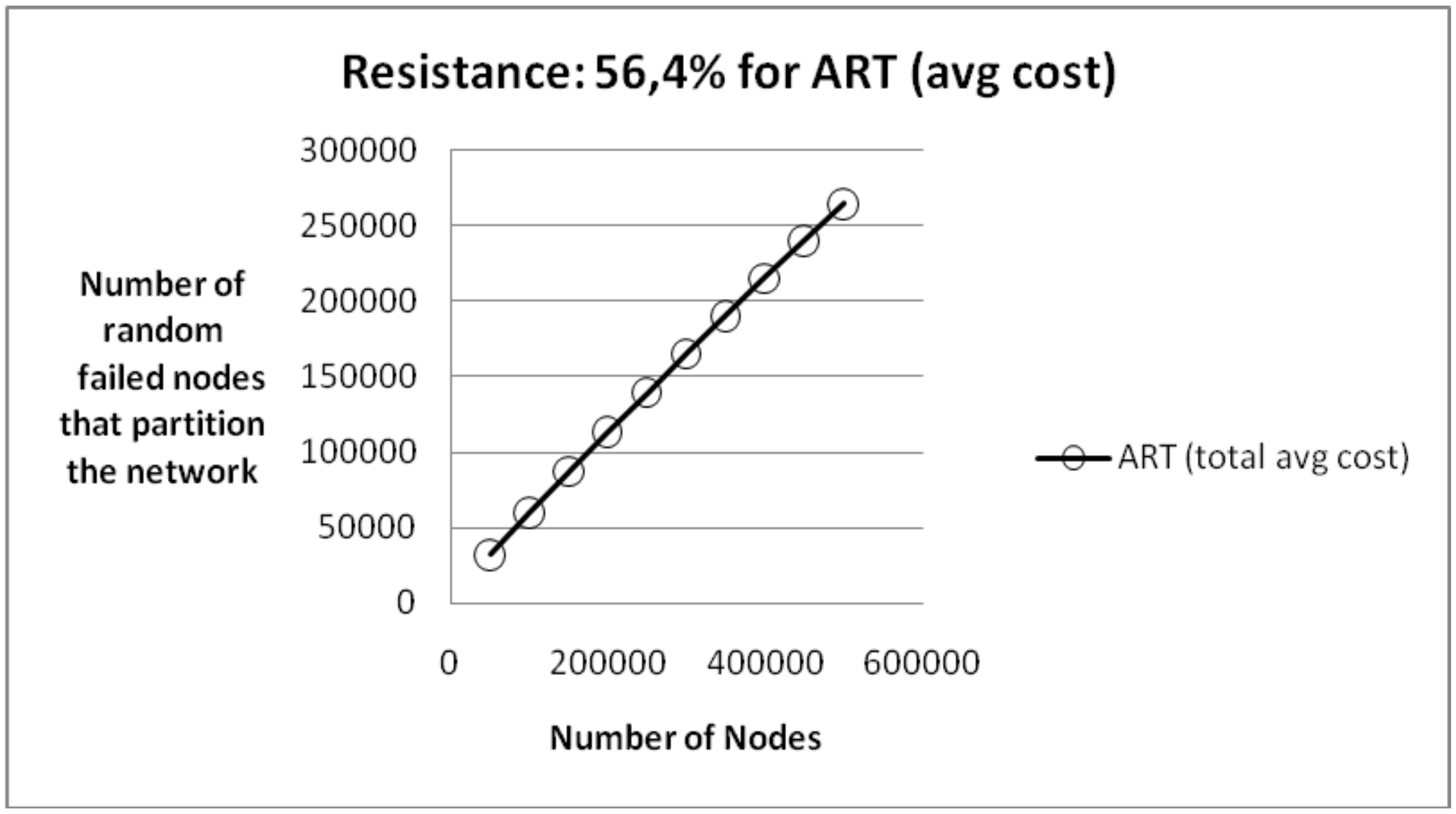

*Figure 13(b). ART Simulations with up to 600K P2P nodes: Resistance percentage avg cost measure*

Figures 14 show a number of more metrics for Chord, and ART. Next, we will describe more metrics for a variety of p2p protocols at the PlanetLab.

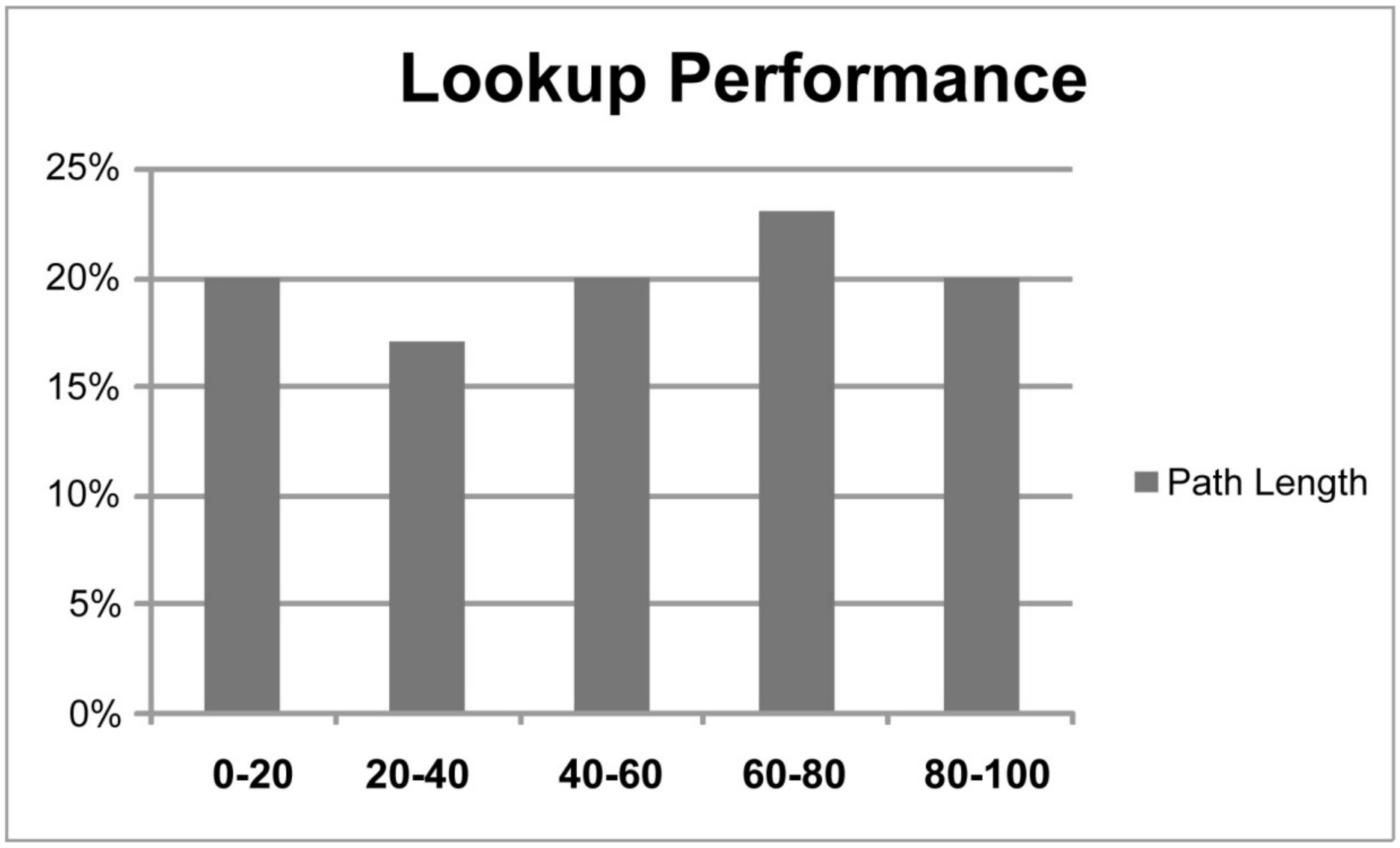

*Figure 14(a). More metrics for Chord path-length: Chord 10K network*

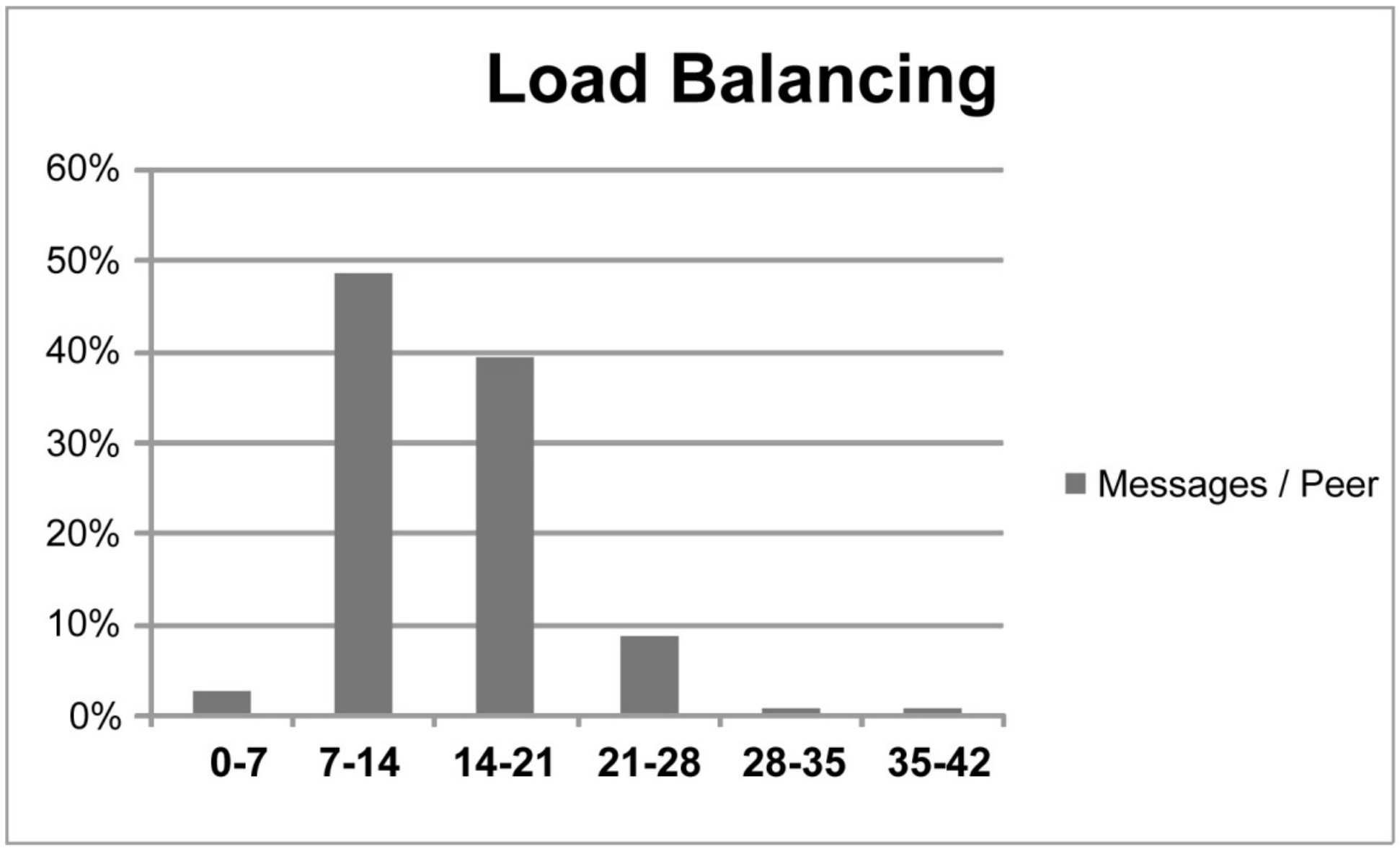

*Figure 14(b). More metrics for ART: Load Balancing (msg/node) Cost Measure for ART 10K network*

# D-P2P-Si*m*+ at the PlanetLab

PlanetLab is a distributed research networks infrastructure comprising 1091 computers, which are found in 505 different research locations. Each network participant allocates computers and is given the possibility to use resources from all the network to implement large scale experiments. The need for transparent transfer of experimental configuration and evaluation into large- scale infrastructures with limited changes is a common vision for researchers and research developers. D-P2P-Si*m*+ has been verified to be easy to setup and execute at the PlanetLab network using the same experimental configuration as that inside a research lab. In order to avoid involvement of PlanetLab computers that are out of reach or overloaded the CoMon tool can be utilized that it provides statistics for PlanetLab available resources, both at nodes and slice level.

In order to test and experiment on a large number of computers, a tool that will connect and execute command in parallel is needed. For this PlanetLab *tools of management slice* are employed (Figure 15). In the experimental simulation scenarios, 50 PlanetLab points were selected, based on their statistics according to the CoMon tool, so that they would not face severe problems of network interconnection. In each evaluation, node failure and recovery scenarios have been evaluated as well as node departures and substitution in order to test fault tolerance and robustness of each protocol tested. During experimentation at the PlanetLab, we observed that execution times are an order of magnitude larger than those required in our lab.

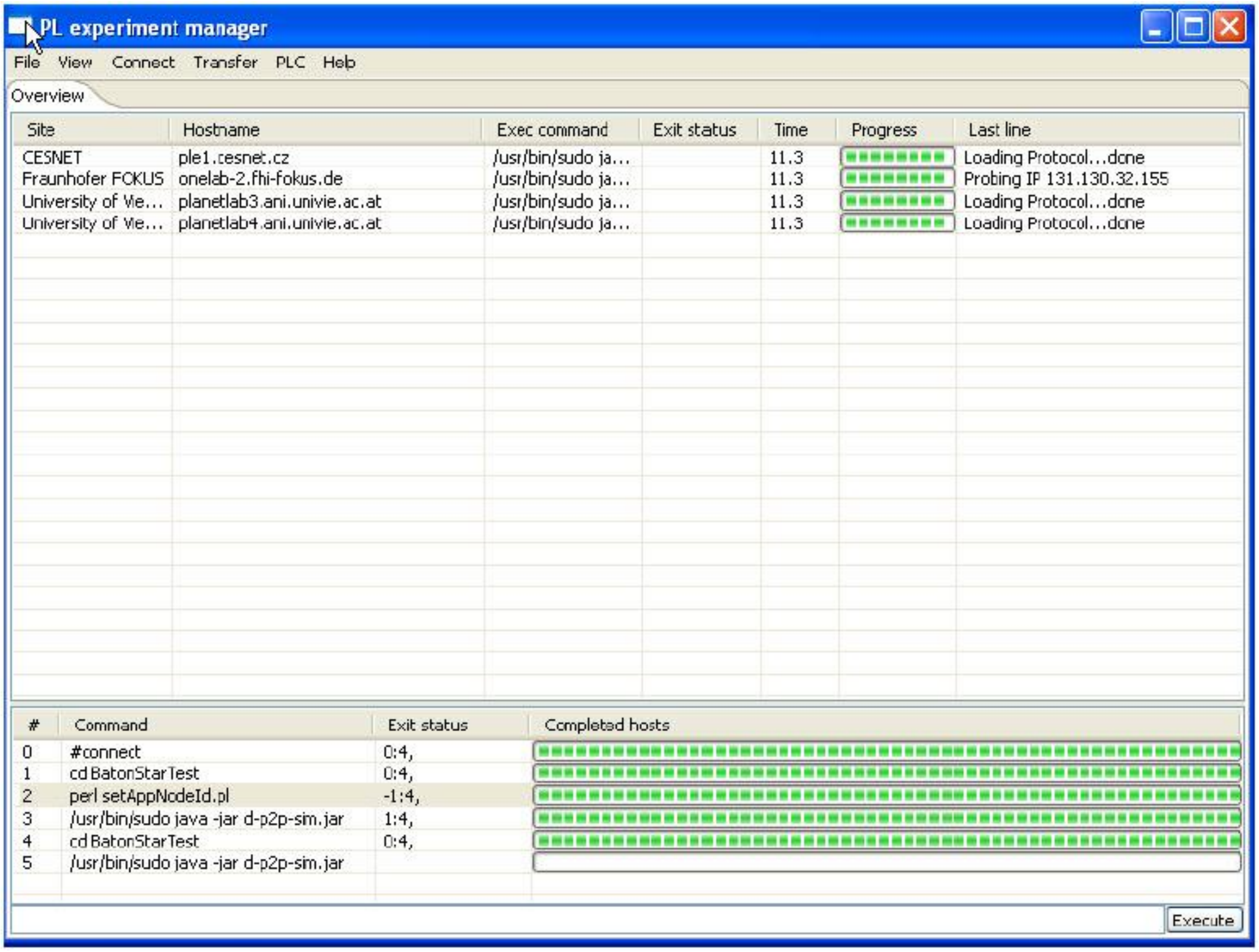

*Figure 15. Executing D-P2P-Sim+ on PlanetLab Network*

There exist cases where 6,000 P2P nodes in PlanetLab take approximately 7.5 hours to return results. Slices of the PlanetLab are often over-loaded and large-scale experiments face enormous delays. Moreover, network communication among PlanetLab slices is an additional overhead that has to be taken into consideration when planning experiments. However, Planetlab is useful in order to verify that an engine under research is possible to work even under very large communication obstacles and most importantly in wide area networks. In figure 15, one notices how D-P2P-Sim$^{+}$ is executed progressively for Baton* P2P at the planetlab nodes.

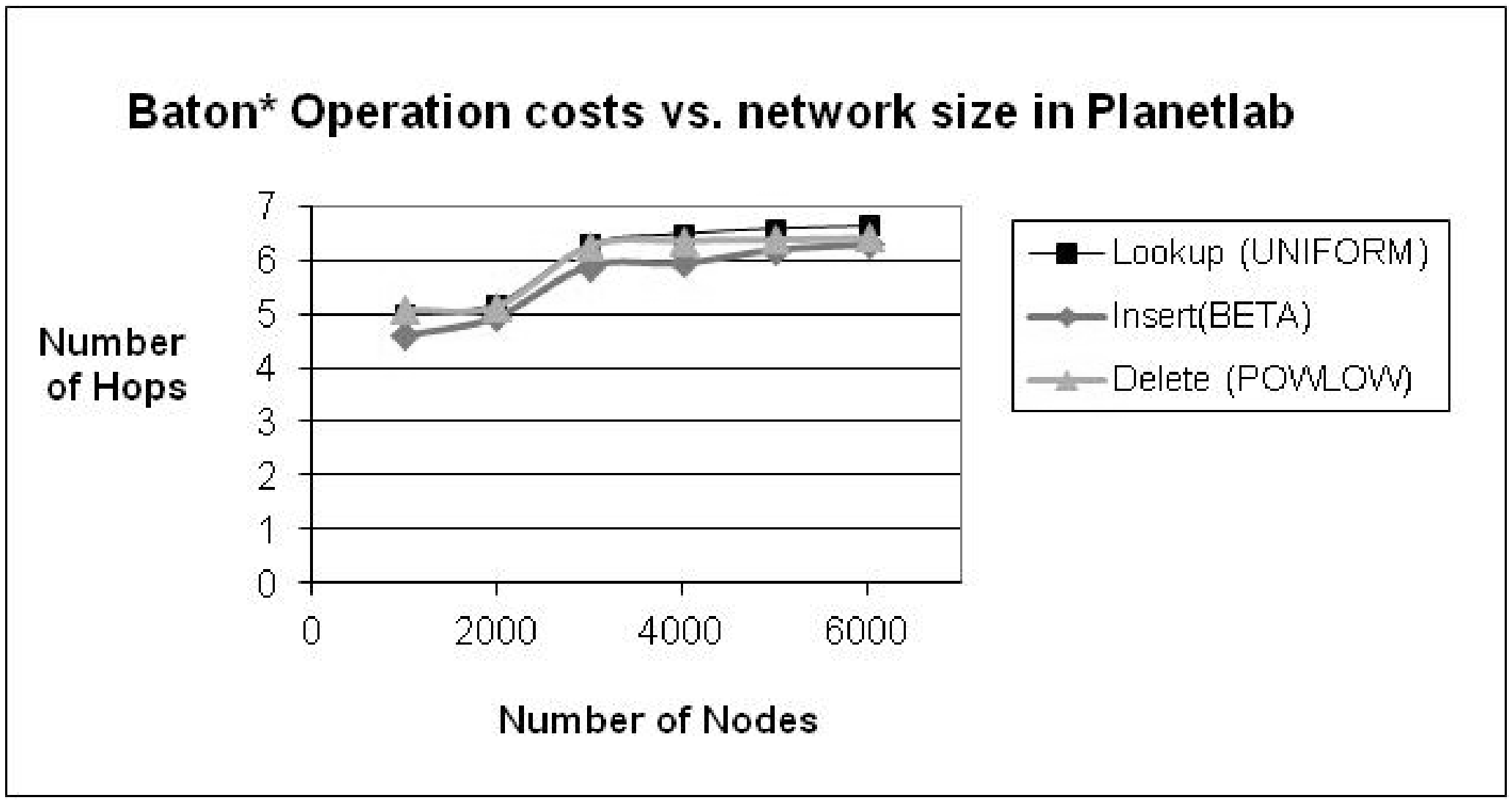

*Figure16. More metrics for BATON*: Baton* Operation costs vs. network size on Planetlab*

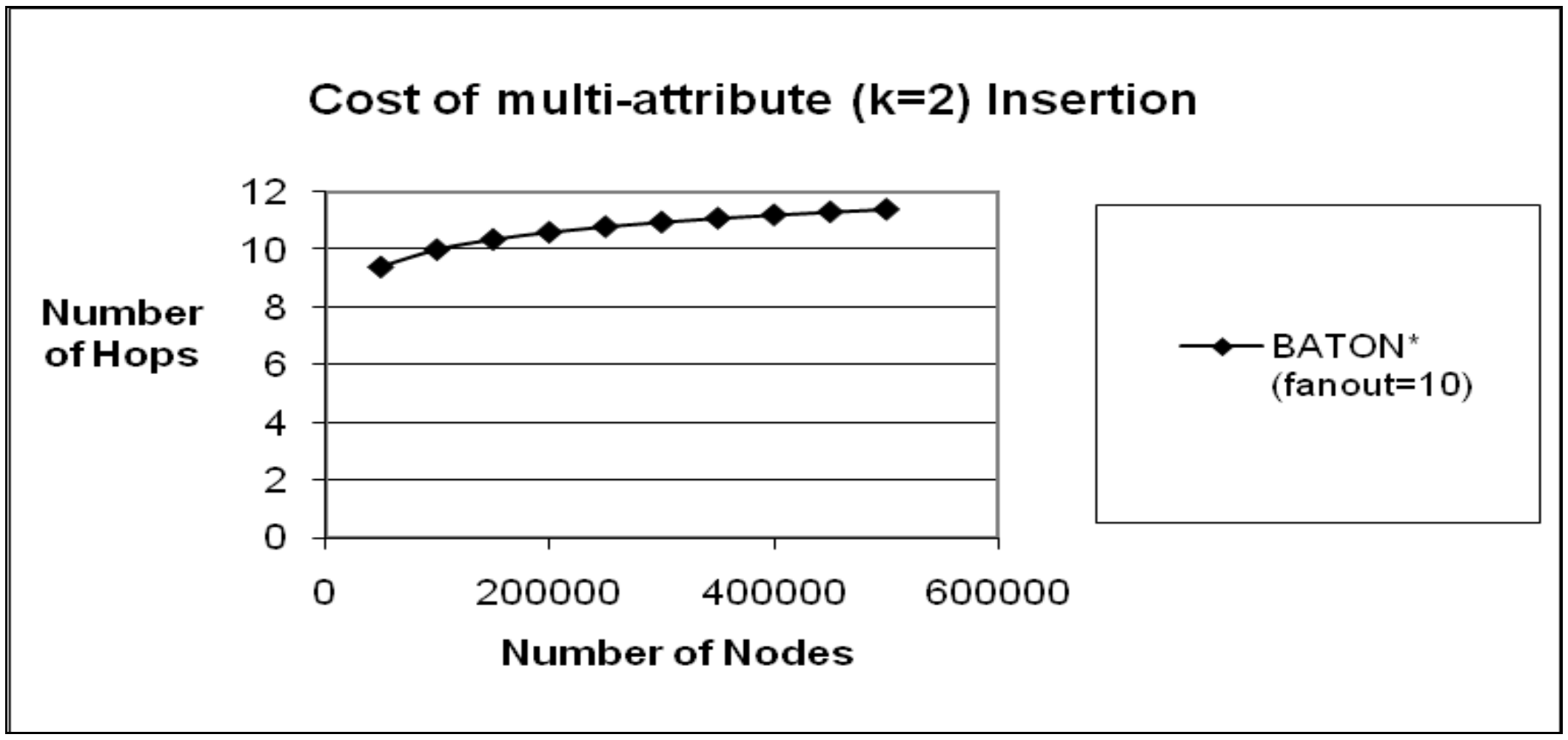

*Figure 17(a). More metrics for BATON* with up to 600K P2P nodes: 2-Dimensional Insertion Cost Measure on Planetlab (Avg Cost in a variety of random distributions)*

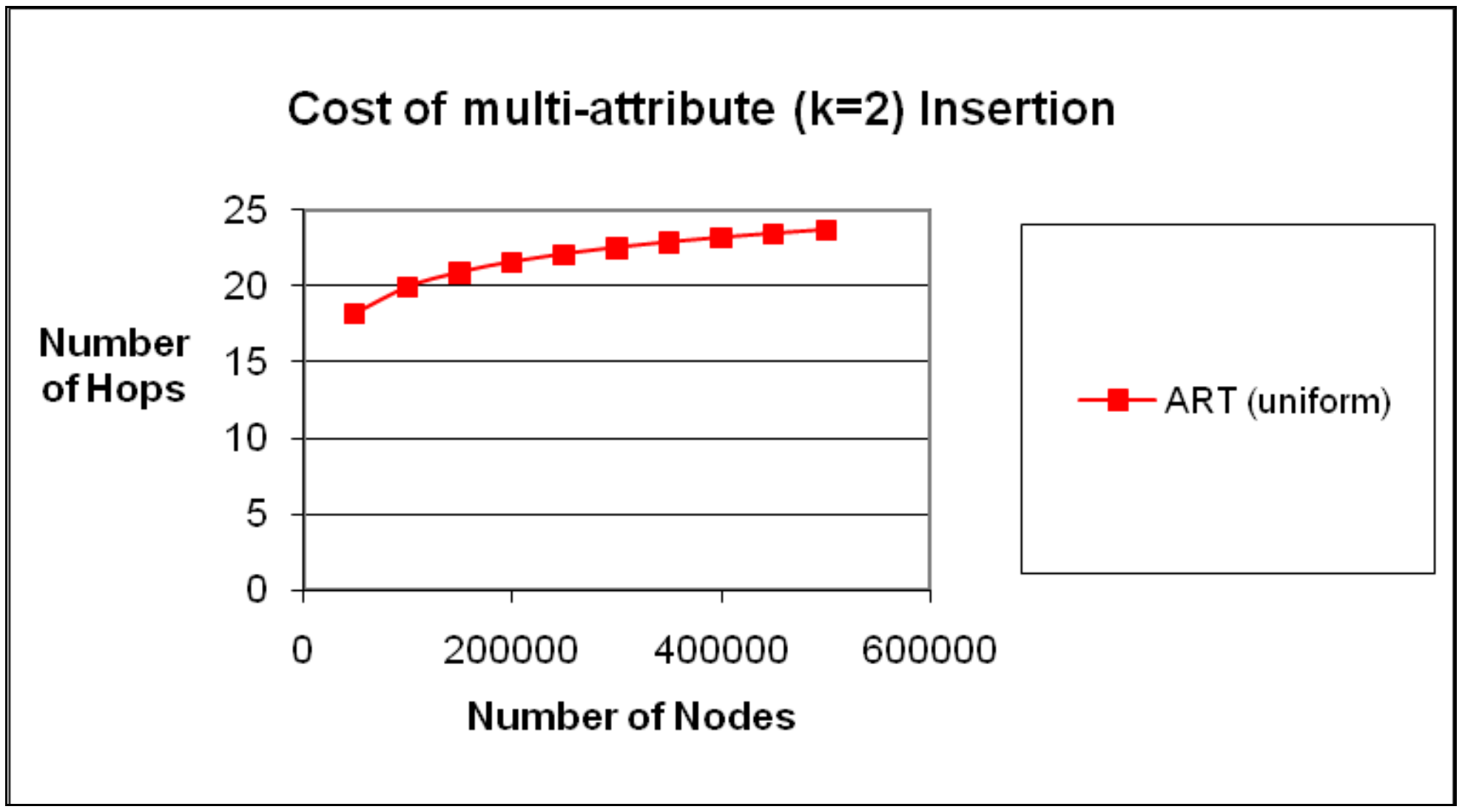

*Figure 17(b). More metrics for ART with up to 600K P2P nodes: 2-Dimensional Insertion Cost Measure on Planetlab for uniform distribution*

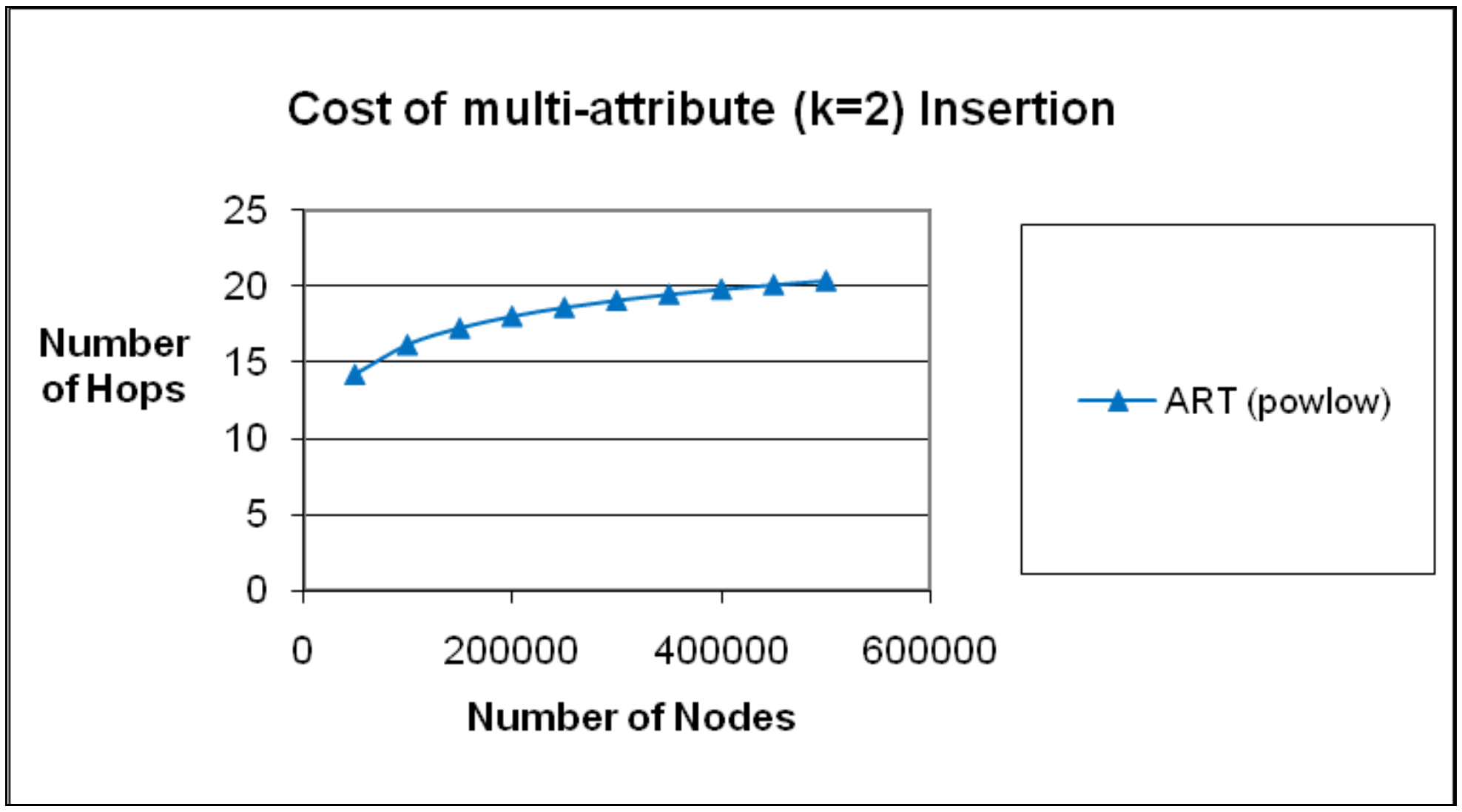

*Figure 17(c). More metrics for ART with up to 600K P2P nodes: 2-Dimensional Insertion Cost Measure on Planetlab for powlow distribution*

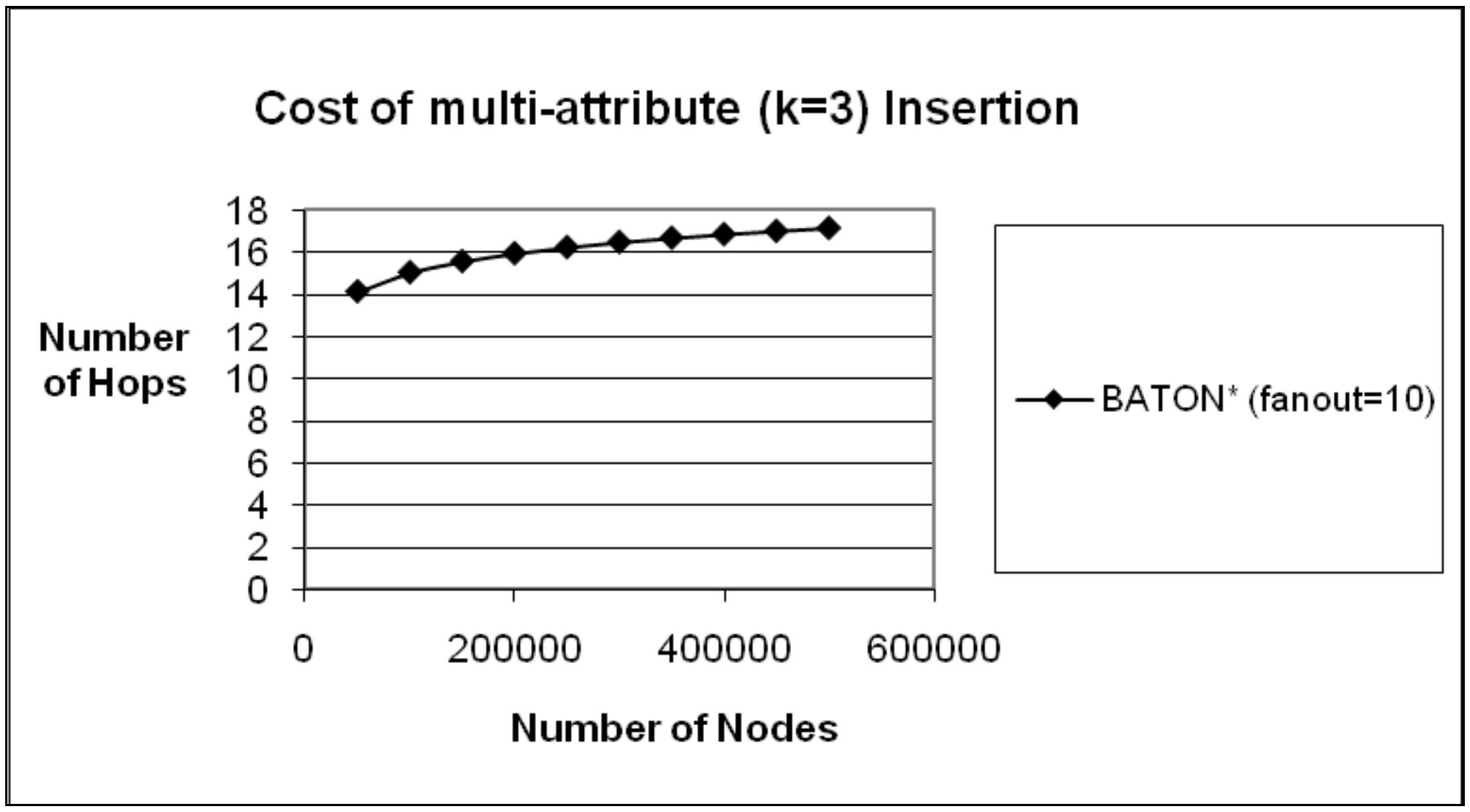

*Figure 18(a). More metrics for BATON* with up to 600K P2P nodes: 3-Dimensional Insertion Cost Measure on Planetlab (Avg Cost in a variety of random distributions)*

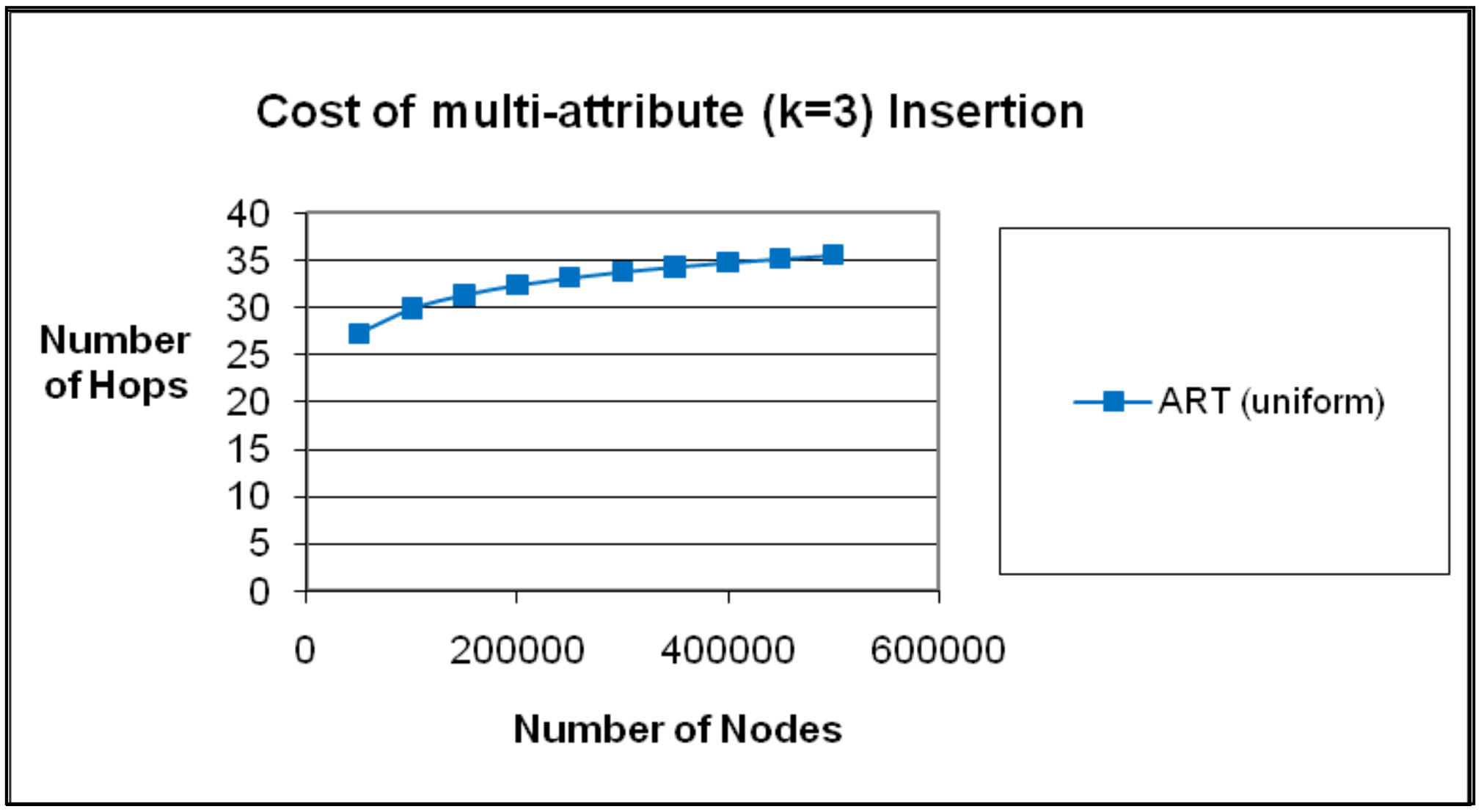

*Figure 18(b). More metrics for ART with up to 600K P2P nodes: 3-Dimensional Insertion Cost Measure on Planetlab for uniform distribution*

Figure 16 depicts cost measurements for a variety of operations (search, insert, delete) each of which is executed at the PlanetLab for a variety of input distributions. Figures 17 and 18 depict multi-dimensional (2 and 3 dimensions respectively) insertion cost measurements in PlanetLab for Baton* and ART.

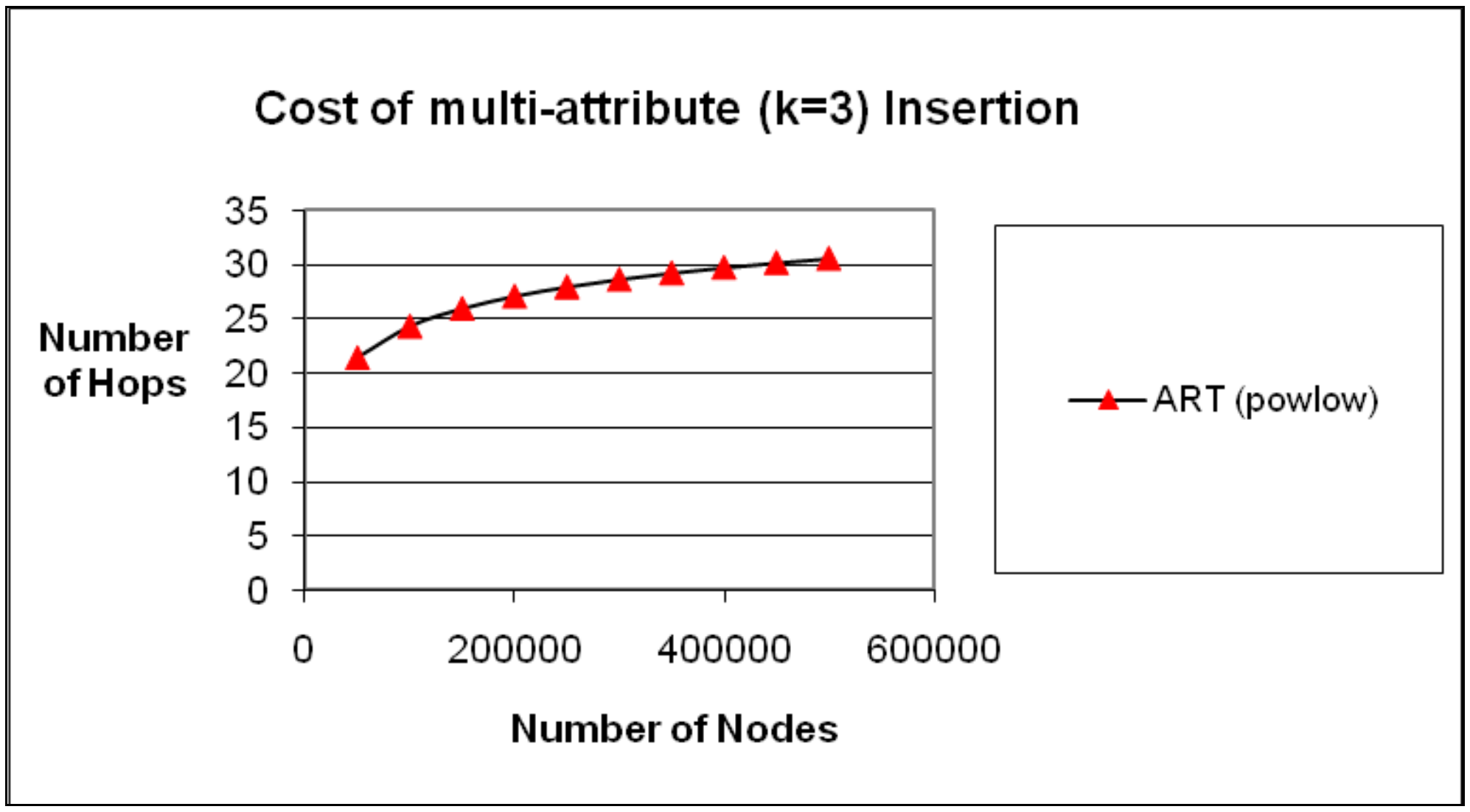

*Figure 18(c). More metrics for ART with up to 600K P2P nodes: 3-Dimensional Insertion Cost Measure on Planetlab for powlow distribution*

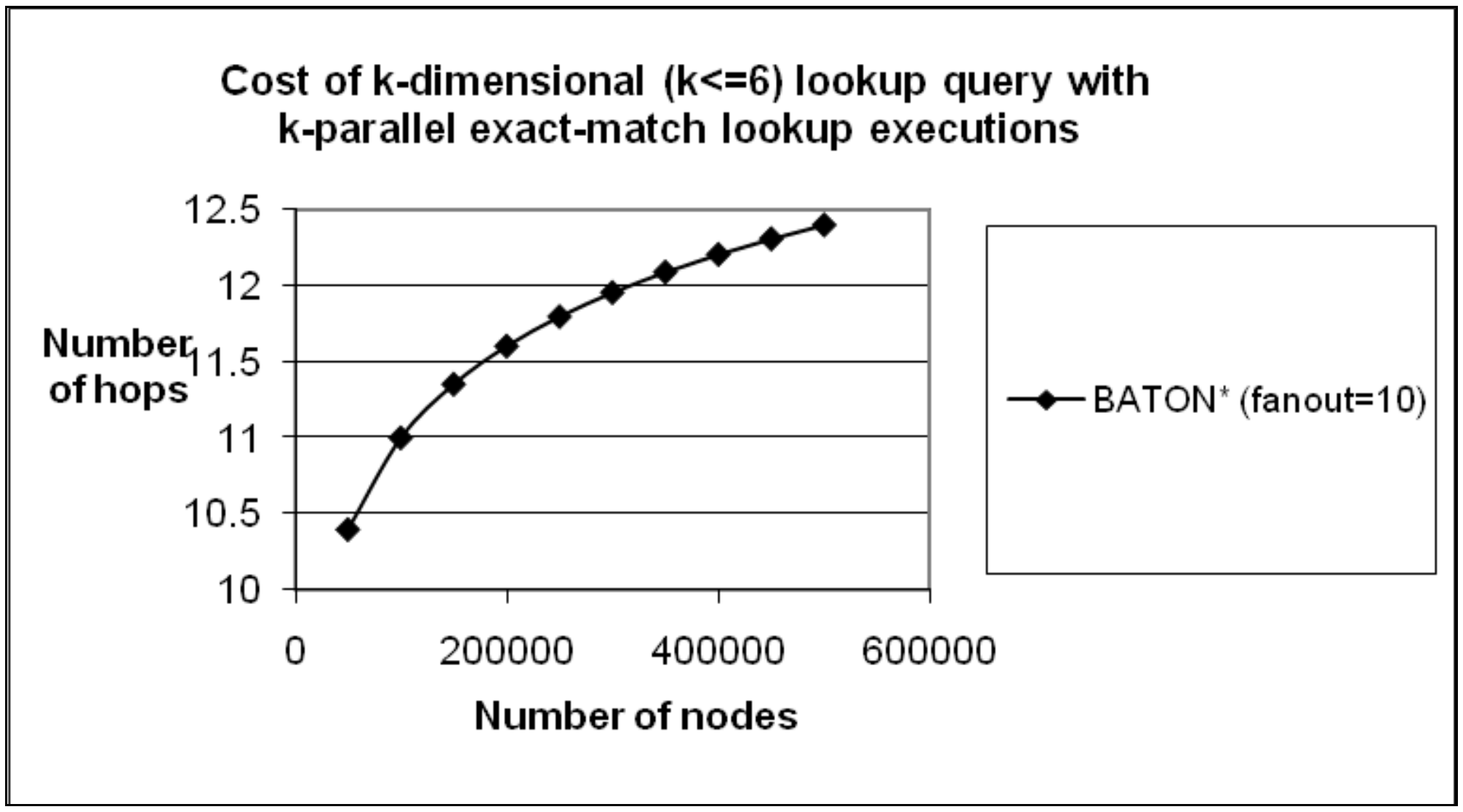

*Figure 19(a). More metrics for BATON* with up to 600K P2P nodes: Multi-Dimensional Lookup Cost Measure on Planetlab (Avg Cost in a variety of random distributions)*

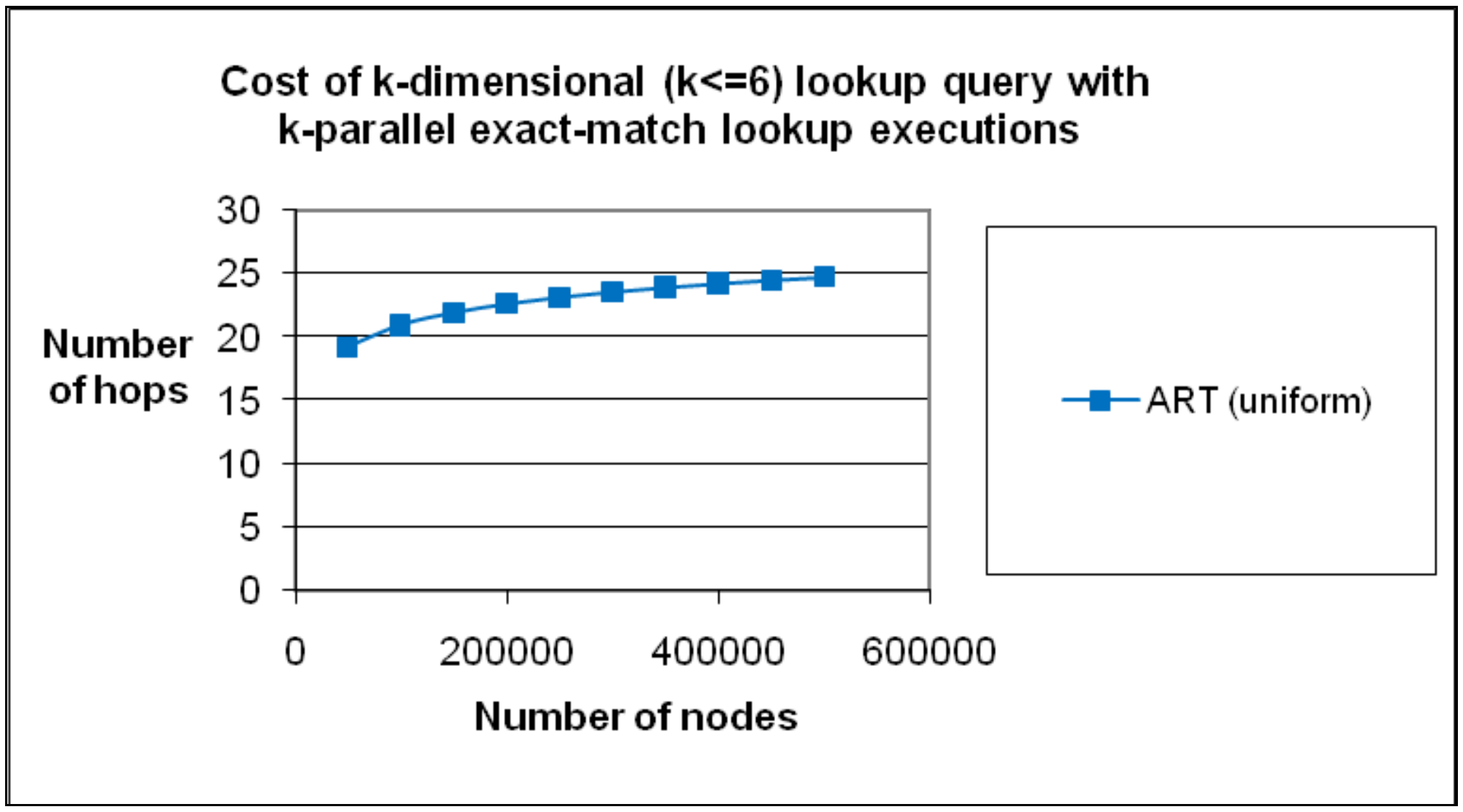

*Figure 19(b). More metrics for ART with up to 600K P2P nodes: Multi-Dimensional Lookup Cost Measure on Planetlab for uniform distribution*

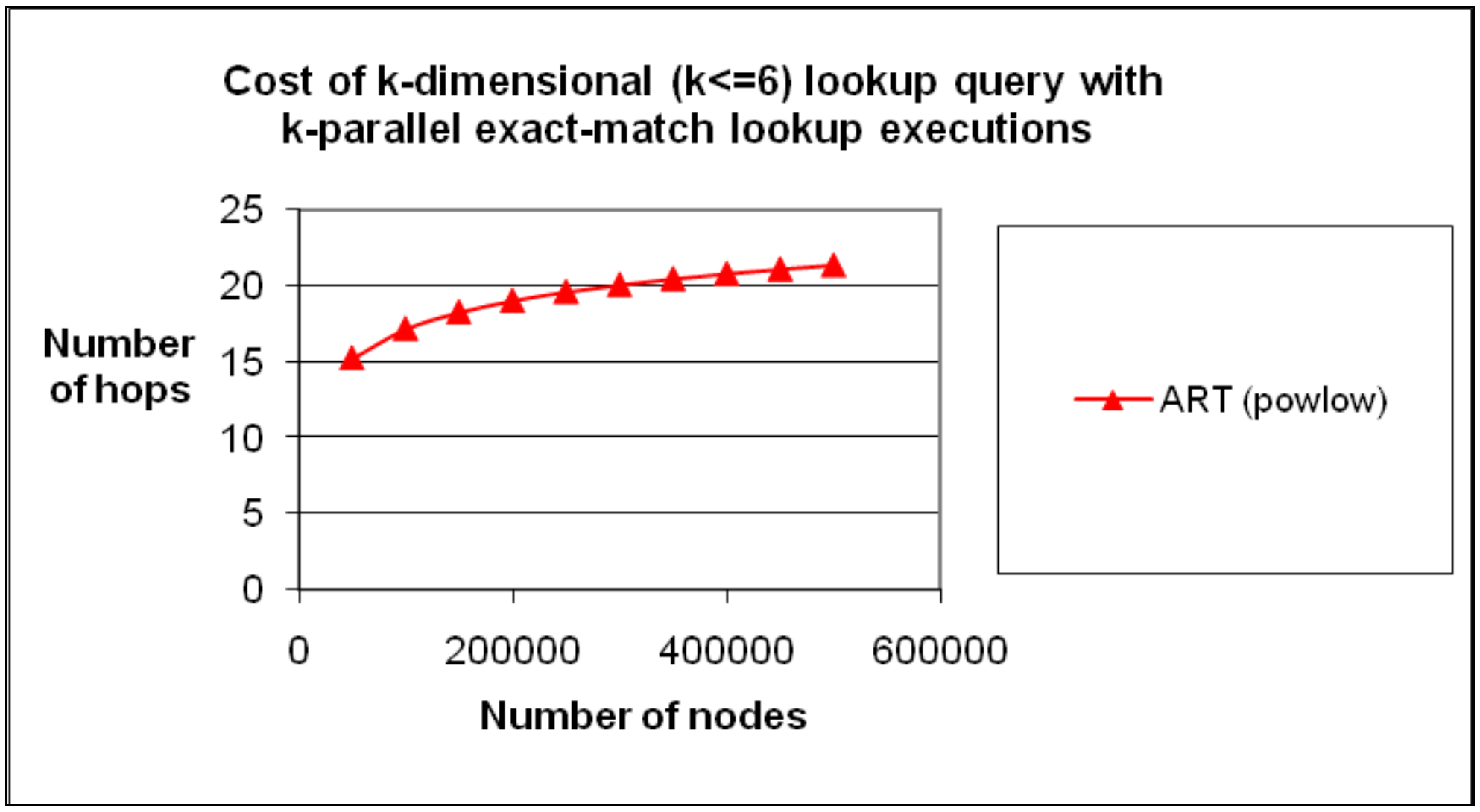

*Figure 19(c). More metrics for ART with up to 600K P2P nodes: Multi-Dimensional Lookup Cost Measure on Planetlab for powlow distribution*

Figures 19 depict multi-dimensional lookup (exact-match) cost measurements in PlanetLab for Baton* and ART for up to six(6) dimensions.

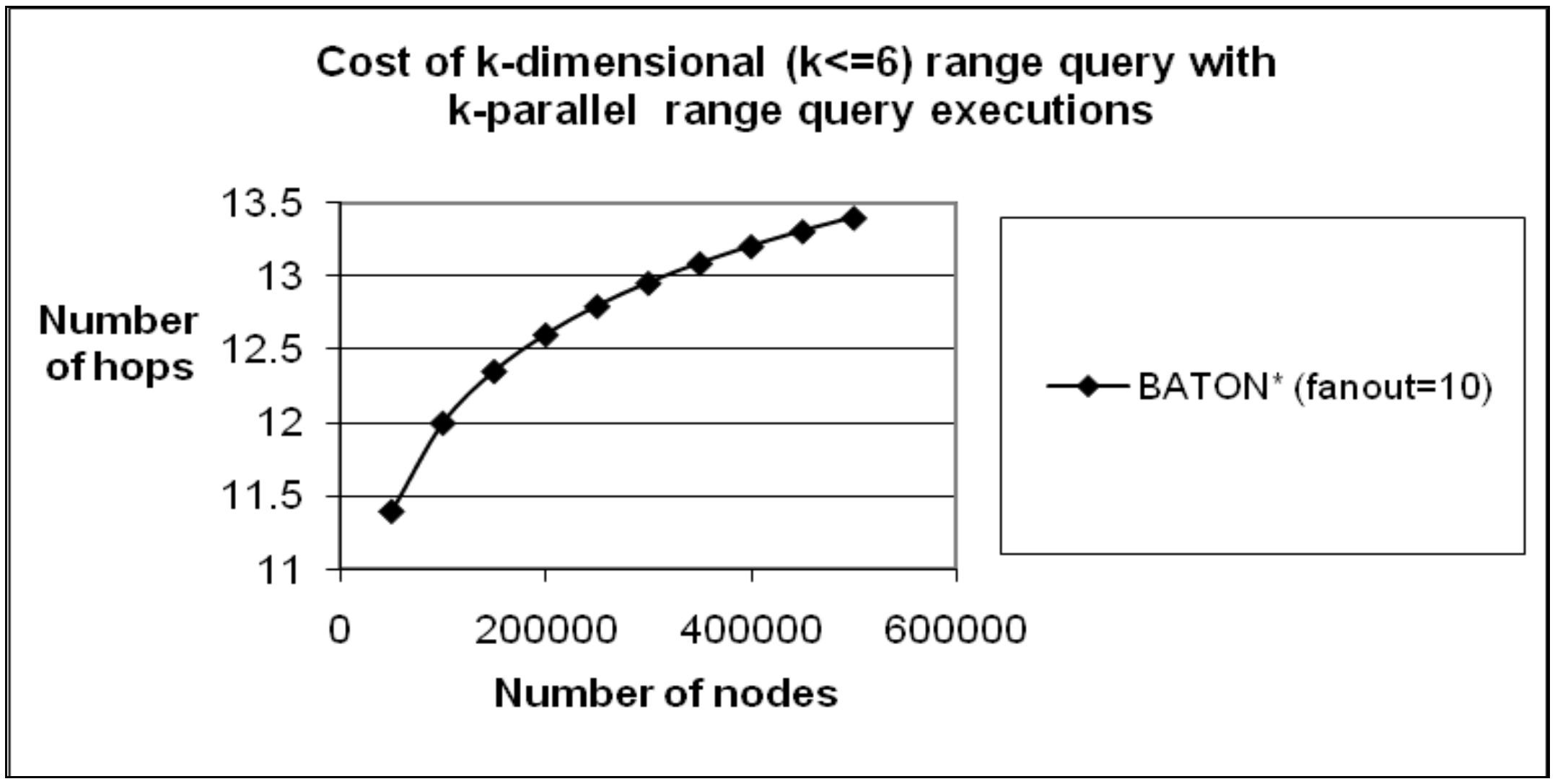

*Figure 20(a). More metrics for BATON* with up to 600K P2P nodes: Multi-Dimensional Range Query Cost Measure on Planetlab (Avg Cost in a variety of random distributions)*

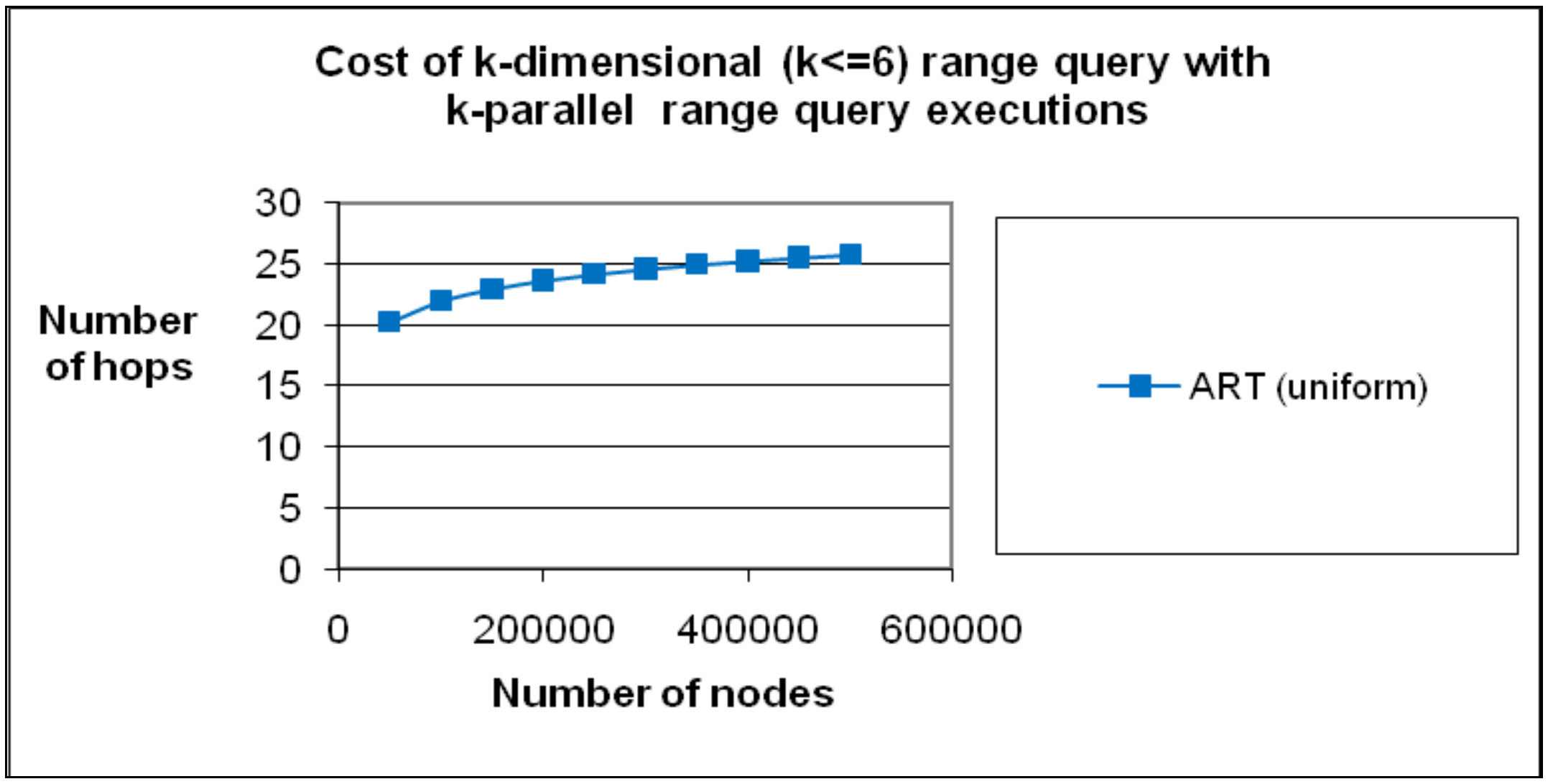

*Figure 20(b). More metrics for ART with up to 600K P2P nodes: Multi-Dimensional Range Query Cost Measure on Planetlab for uniform distribution*

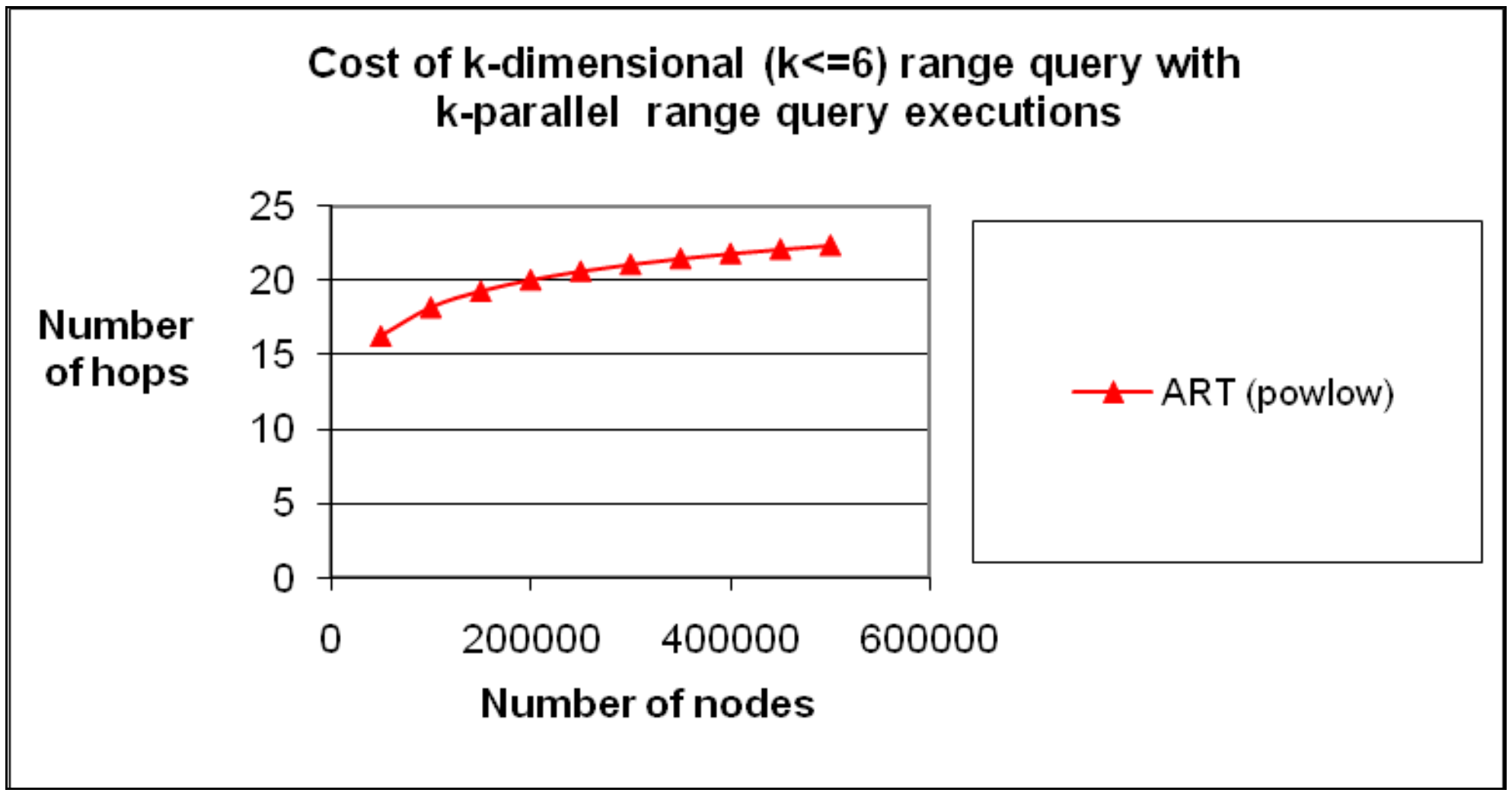

*Figure 20(c). More metrics for ART with up to 600K P2P nodes: Multi-Dimensional Range Query Cost Measure on Planetlab for powlow distribution*

Figures 20 depict multi-dimensional range-query cost measurements in PlanetLab for Baton* and ART for up to six (6) dimensions.

## Accuracy Aspects of D-P2P-Sim+

According to (Basu et al 2013) , if P2P simulation is able to accurately simulate real P2P networks, it can give the ability to design, implement and evaluate new technologies in a controlled environment, while allowing for the collection of statistics with greater ease, compared to deployment with real users. In general, it is practically impossible to capture the full complexity of the Internet in any controlled environment. To achieve even higher accuracy when using the proposed simulator, we have designed, deployed, experimented and verified the results presented above over the top real world test bed environment known as the global research network PlanetLab (https://www.planet-lab.org). PlanetLab test bed consists of over 1000 computers situated at participating sites across the globe. PlanetLab has already several years' history and it continuously expands with new Universities and Research Centers joining its infrastructure.

However, while such an infrastructure provides valuable real-world data in terms of network latency, the disadvantage is that a researcher has to depend on the machines with limited and sometimes unexpected availability and to make appropriate configuration steps each time on PlanetLab in order to reproduce results. This can be a tedious task since the machines belong in different administrative domains, each with its own support staff and hardware/software upgrade schedule. Therefore, tracking issues and problems across such a large number of real computers is a challenging and daunting task. To avoid all this overhead and allow the researcher to focus on experimental measurements, the proposed solution provides a useful approach-strategy to smooth and overcome difficulties.

First, D-P2P-Sim+ enables the researcher to perform real-world simulation using multiple nodes on small and medium numbers of physical nodes. Physical nodes in this case are computers that the researcher may have available locally in a laboratory or at collaborating laboratories. Having simulation in such laboratory environments allow the researcher to schedule more easily and guarantee their availability for simulation purposes than in the large scale network of PlanetLab.

A key feature of D-P2P-Sim+ is that it may simulate multiple P2P nodes at each physical node used in local or wide area networks. Therefore, the researcher is able to setup simulation experiments with large numbers of nodes P2P networks in his/her laboratory efficiently, before deployment and exposure to the vagaries of the real world.

D-P2P-Sim+ allows for the representation of nodes (which simulate peers) with incoming and outgoing message queues. Over time steps these message queues are processed, passing messages between peer outgoing and incoming queues, simulating the communication in a P2P network. The time step length is possible to be configured accordingly at each node separately or evenly across all nodes in the experimental setup in order depict appropriate end-to-end delays as determined by real life network topologies and prevailing network patterns (Lin et al, 2005). All nodes are aware of their neighbors, and therefore send and receive messages to their neighbors via their message queues (Vahdat et al 2002). Moreover, D-P2P-Sim+ is possible to simulate node failure and recovery strategies and experimental scenarios to evaluate the behavior of the P2P algorithm under research in such cases.

Next, the novel D-P2P-Sim+ is ready to be used through the PlanetLab network, whenever the researcher needs additional physical nodes to experiment with a simulation for example to insert additional nodes in the P2P of to utilize and test a specific network topology of previously unknown characteristics. Please note that, we have verified that experimental results from PlanetLab about P2P networks do follow the produced results at a single laboratory and multiple collaborating laboratories as shown at the figures above. We show that a limited number of modern typical PCs can accurately simulate network latency and failures of multi-millions of nodes in a laboratory environment.

Based on the above principles, we presented the novel D-P2P-Sim+, which has the ability to be appropriately configured in both, small-scale LAN network laboratories and if needed in large scale a real world test bed environment such as PlanetLab. Evaluation has shown that PlanetLab experimental results hopefully verified a “beautiful” reproduction of local lab results with a limited number of PCs only. The latter proves that our simulator does capture real-world performance characteristics accurately with appropriate configuration.

Overall, we have designed D-P2P-Sim+ to include: i) features to allow reducing the complexity of the simulated topology, ii) options to allow multiple virtual nodes operating onto a single physical machine, and iii) introducing synthetic background traffic to dynamically change network characteristics and to simulate network failures.

## Conclusions

D-P2P-Si*m*+ is presented as a novel framework to support simulation with multi million nodes storing large amount of data collected in a variety of IoT applications. The simulator allows experimenting when designing new protocols and overlays for data stored in P2P systems, thus identifying the ideal scalable solution for efficient data management and satisfaction of SLAs in the actual environment. Contrary to the rest of P2P simulators, it supports the deployment of any distributed indexing and processing engine.

D-P2P-Si*m*+ allows wide scale P2P network experimentation in the laboratory environment. Moreover it includes a robust framework to work easily on failure and recovery scenarios for the networks under design or research. Its robust statistics are supported and all operations are available through a simple GUI developed using Java tools.

We have shown that D-P2P-Si*m*+ is an ideal platform for storing and querying the collected data in a fully customized way. Experimental evaluation in a real world test bed, such as PlanetLab, verified accurately the theoretic analysis of a variety of p2p protocols for up to 2 M nodes. Additional technical details, architectural blueprints and multiple snapshots are available online at the framework's web-page.

**Acknowledgments:** We would like to thank Prof. E. Kafeza for helpful comments and fruitful discussions..

# References

A. S. report on Skype usage. (2011). Retrieved from http://news.softpedia.com/news/skype usage-surges-as-27-million-people-chat-simultaneously-177523.shtml

Atzori, L., Iera, A., & Morabito, G. (2010). The Internet of Things: A survey. *Computer Networks*, *54*(15), 2787-2805., doi: 10.1016/j.comnet.2010.05.010

Basu, A., Fleming, S., Stanier, J., Naicken, S., Wakeman, I., Gurbani, V.K., 2013. “The state of peer-to-peer network simulators”, ACM Computing Surveys ACM Comput. Surv. 45, 4, Article 46 (August 2013), 25 pages

Bertino, E., Guerrini, G. & Mesiti, M. (2004). A matching algorithm for measuring the structural similarity between an xml document and a dtd and its applications. *Information Systems*, *29*(1),23–46

Bobelin, L., Legrand, A, González Márquez, D., Navarro, P., Quinson, M., Suter, F., & Thiery,C., 2012. *Scalable Multi-purpose Network Representation for Large Scale Distributed System Simulation*. 2012 12th IEEE/ACM International Symposium on Cluster, Cloud and Grid Computing (CCGRID '12). IEEE Computer Society, Washington, DC, USA, 220-227

Brusilovsky, P., & Maybury, T. (2002). From adaptive hypermedia to the adaptive web. *Communications of the ACM Journal*, *45*(5), 31–33

Brusilovsky, P., Kobsa, A., & Nejdl, W. (2007). The adaptive web: methods and strategies of web personalization. *4321 LNCS*, Springer

Casanova, H., Legrand, A., & Quinson, M., 2008. *SimGrid: A generic framework for large-scale distributed experiments*. In Proc. of Int. Conf. on Computer Modeling and Simulation, Los Alamitos, CA, USA, 2008. IEEE Comp. Soc., 126–131

Chen, L., Cui, B., & Lu, H. (2011). Constrained skyline query processing against distributed data sites. IEEE Transactions on Knowledge and Data Engineering, 23, 204–217.

Cowie, J., Liu, H., Liu, J., Nicol, D., & Ogielski, A. (1999). *Towards realistic million-node internet simulation*. In Proceedings of the International Conference on Parallel and Distributed Processing Techniques and Applications, June 28 - Jully 1, Las Vegas, Nevada, USA, 2129–2135

Cuzzocrea, A. (2006). Combining multidimensional user models and knowledge representation and management techniques for making web services knowledge-aware. *Web Intelligence and Agent Systems: An international journal, 4*, 289–312

Jagadish, H. V., Ooi, B. C., Tan, K.-L., Vu, Q. H., & Zhang, R. (2006). *Speeding up search in peer-to-peer networks with a multi-way tree structure*. In Proceedings of the ACM SIGMOD International Conference on Management of Data, Chicago, Illinois, USA, June 27-29, 1–12

Lai, K-C., Yu, Y-F., 2012. A scalable multi-attribute hybrid overlay for range queries on the cloud, *Information Systems Frontiers*, Springer, September 2012, Volume 14, Issue 4, pp 895-908

Lin, S., Pan, A., Guo, R., & Zhang, Z., 2005. 13th International Symposium on Modeling, Analysis, and Simulation of Computer and Telecommunication Systems, USA, pp. 415-424

Mayer, T., Coquil, D., Schoernich, C., & Kosch., H., 2012. *RCourse: a robustness benchmarking suite for publish/subscribe overlay simulations with Peersim*. In Proceedings of the First Workshop on P2P and Dependability (P2P-Dep '12). ACM, New York, NY, USA, , Article 3 , 6 pages

Montresor, A., & Jelasity, M., 2009. *PeerSim: A scalable P2P simulator*. In Proc. of the 9th Int. Conference on Peer-to-Peer (P2P'09), pages 99-100, Sept. 2009

Miorandi, D., Sicari, S., De Pellegrini, F., & Chlamtac, I. (2012). Internet of things: Vision, applications and research challenges. *Ad Hoc Networks, 10*(7), 1497-1516. doi: 10.1016/j.adhoc.2012.02.016

Naicken, S., Livingston, B., Basu, A., Rodhetbhai, S., Wakeman, I., & Chalmers, D. (2007). The state of peer-to-peer simulators and simulations. *SIGCOMM Comput. Commun. Rev., 37*(2), 95–98

Personalization reports. (n.d.). Retrieved from http://www.choicestream.com/who/news.php

Quinson, M., Rosa, C., & Thiery., C., 2012. *Parallel Simulation of Peer-to-Peer Systems*. In Proceedings of the 2012 12th IEEE/ACM International Symposium on Cluster, Cloud and Grid Computing (ccgrid 2012) (CCGRID '12). IEEE Computer Society, Washington, DC, USA, 668-675

Shmueli-Scheuer, M., Roitman, H., Carmel, D., Mass, Y., & Konopnicki, D. (2010). *Extracting user profiles from large scale data*. In Proceedings of the 2010 Workshop on Massive Data Analytics on the Cloud

Sioutas, S. (2008). NBDT: An efficient p2p indexing scheme for web service discovery. *Journal of Web Engineering and Technologies, 4*(1), 95-113

Sioutas, S., Papaloukopoulos, G., Sakkopoulos, E., Tsichlas, K., & Manolopoulos, Y. (2009). *A novel distributed p2p simulator architecture: D-p2p-sim*. In Proceedings of the 18th ACM Conference on Information and Knowledge Management, CIKM 2009, Hong Kong, China, November 2-6, 2069–2070

Sioutas, S., Papaloukopoulos, G., Sakkopoulos, E., Tsichlas, K., Manolopoulos, Y., & Triantafillou, P. (2010). *Brief announcement: Art–sub-logarithmic decentralized range query processing with probabilistic guarantees*. In ACM PODC 2010, 118-119

Teng, H.Y., Lin, C.N., & Hwang, R-H., 2013. A self-similar super-peer overlay construction scheme for super large-scale P2P applications, *Information Systems Frontiers*, Springer, September 2013, DOI: 10.1007/s10796-013-9456-3

Vahdat, A., Yocum, K., Walsh, K. Mahadevan, P., Kostic, D., Chase, J. & Becker, D., 2002. “Scalability and Accuracy in a Large-Scale Network Emulator”, in ACM SIGOPS Operating Systems Review - OSDI '02: Proceedings of the 5th symposium on Operating systems design and implementation, Vol. 36 No. SI, pp. 271-284

Wang, J., Sharman, R., & Ramesh, R. (2008). Shared content management in replicated web systems: A design framework using problem decomposition. *IEEE Trans on Systems, Man and Cybernetics PART C, 38*(1), 110–124

Windows Live Messenger: A short history. (2010).
Retrived from http://windowsteamblog.com/windows_live/b/windowslive/archive/2010/02/09/windows-live-messenger-a-short-history.aspx

Zhang, A. N., Goh, M., & Meng, F. (2011). Conceptual modelling for supply chain inventory visibility. *International Journal of Production Economics, 133*(2), 578-585. doi: 10.1016/j.ijpe.2011.03.003